\documentclass[12pt]{article}

\def\ltwid{\mathrel{\raise.3ex\hbox{$<$\kern-.75em\lower1ex\hbox{$\sim$}}}}
\def\comp{{\rm C}\llap{\vrule height7.1pt width1pt depth-.4pt\phantom t}}

\begin{document}

\begin{titlepage}
\begin{flushright}
CRETE-08-11 \\ UFIFT-QG-08-05
\end{flushright}

\vspace{0.5cm}

\begin{center}
\bf{A SIMPLIFIED QUANTUM GRAVITATIONAL MODEL OF INFLATION}
\end{center}

\vspace{0.3cm}

\begin{center}
N. C. Tsamis$^{\dagger}$
\end{center}
\begin{center}
\it{Department of Physics, University of Crete \\
GR-710 03 Heraklion, HELLAS.}
\end{center}

\vspace{0.2cm}

\begin{center}
R. P. Woodard$^{\ddagger}$
\end{center}
\begin{center}
\it{Department of Physics, University of Florida \\
Gainesville, FL 32611, UNITED STATES.}
\end{center}

\vspace{0.3cm}

\begin{center}
ABSTRACT
\end{center}
\hspace{0.3cm}
Inflationary quantum gravity simplifies drastically in the leading logarithm 
approximation. We show that the only counterterm which contributes in this
limit is the 1-loop renormalization of the cosmological constant. We go
further to make a simplifying assumption about the operator dynamics at 
leading logarithm order. This assumption is explicitly implemented at 1- 
and 2-loop orders, and we describe how it can be implemented 
nonperturbatively. We also compute the expectation value of an invariant 
observable designed to quantify the quantum gravitational back-reaction on 
inflation. Although our dynamical assumption may not prove to be completely 
correct, it does have the right time dependence, it can naturally produce  
primordial perturbations of the right strength, and it illustrates how a 
rigorous application of the leading logarithm approximation might work in 
quantum gravity. It also serves as a partial test of the ``null hypothesis'' 
that there are no significant effects from infrared gravitons.

\vspace{0.3cm}

\begin{flushleft}
\end{flushleft}

\vspace{0.1cm}

\begin{flushleft}
$^{\dagger}$ e-mail: tsamis@physics.uoc.gr \\
$^{\ddagger}$ e-mail: woodard@phys.ufl.edu
\end{flushleft}

\end{titlepage}

\section{Introduction}

There are profound differences between quantum gravity with a positive 
cosmological constant and its flat space version . Many of these
differences are associated with the phenomenon of {\it infrared logarithms}. 
These are factors of the number of e-foldings since the onset of inflation 
that show up in the Green's functions of theories which incorporate either 
gravitons or massless, minimally coupled scalars. Their physical origin is 
the continual creation of long wavelength gravitons and scalars during 
inflation, which engenders a slow growth in the corresponding average
field strengths. Infrared logarithms have been seen in pure quantum 
gravity \cite{TW1}, in gravity + fermions \cite{MW1}, in full scalar-driven 
inflation \cite{SW,UT}, in the scalar sector of scalar-driven inflation 
\cite{many}, in $\varphi^4$ theory \cite{phi4}, in scalar QED \cite{SQED} 
and in Yukawa theory \cite{Yukawa,gW2}.

Infrared logarithms introduce a fascinating, secular element into the
usual static results of flat space quantum field theory. For example, 
whereas the virtual gravitons of flat space have no net effect on the 
propagation of a single fermion, the spin-spin coupling with the sea of 
infrared gravitons produced during inflation makes such a fermion behave
as if its field strength were being amplified by a factor of 
$1/\sqrt{Z_2(t)}$ with \cite{MW1}:
\begin{equation}
Z_2(t) = 1 - \frac{17}{4\pi} \, G H^2 
\ln \! \left( \frac{a(t)}{a_{\rm on}} \right)
+ O(G^2) \; . \label{Z2}
\end{equation}
Here $G$ is the Newtonian gravitational constant, $H$ the inflationary 
Hubble parameter and $a_{\rm on}$ represents the scale factor at the onset
of inflation. The dimensionless product $G H^2 \ltwid 10^{-12}$ is bounded 
by the current upper limit on the tensor-to-scalar ratio \cite{WMAP}. Despite 
the minuscule coupling constant, the continued growth of the scale factor
$a(t)$ over a prolonged period of inflation must eventually result in this 
1-loop correction becoming order unity. Because higher loop corrections 
also become order unity at about the same time it is clear that perturbation 
theory breaks down and that one must employ a nonperturbative technique to 
evolve further. 

The eventual breakdown of perturbation theory evident in expression 
(\ref{Z2}) is a general feature of quantum field theories that exhibit 
infrared logarithms. Starobinski\u{\i} has developed a simple stochastic 
formalism \cite{AAS} which has been proved to reproduce the leading 
infrared logarithms of scalar potential models at arbitrary loop order
\cite{TW2}. When the scalar potential is bounded below this technique 
can even be used to evolve past the breakdown of perturbation theory to 
asymptotically late times \cite{SY}. Starobinski\u{\i}'s technique has
recently been generalized to scalar models which involve fields that
do not produce infrared logarithms such as fermions \cite{gW2} and
photons \cite{PTsW}. However, it has not yet been extended to the 
derivative couplings of quantum gravity.

The purpose of this paper is to explore quantum gravity at leading 
logarithm order. 
\footnote{For some interesting alternate approaches, see \cite{AMP,AMM}.}
We have two results, one rigorous and the other qualitative. The rigorous 
result is that ``leading log'' quantum gravity is much better behaved 
than full quantum gravity. The only counterterm that contributes at 
leading logarithm order is the 1-loop renormalization of the cosmological 
constant. This is derived in Section 2.

Our qualitative result consists of a simplifying dynamical assumption,
the {\it Effective Scale Factor Approximation}, under which the leading 
infrared logarithms of quantum gravity can be computed and summed to give 
nonperturbative results. Although this assumption may not be entirely right, 
its qualitative features are consistent with all that is currently known 
about infrared logarithms, and the approximation allows us to explore how 
a rigorous application of Starobinski\u{\i}'s formalism might work in pure 
quantum gravity. 

Both results stand on their own, however, it is interesting to view them
in the context of our long-held suspicion that inflation might be driven
by a positive bare cosmological constant without any scalar inflaton
\cite{TW0,TW1}. An attractive feature of this idea is that it dispenses
with the usual problem of explaining why the cosmological constant is so
small by assuming that $\Lambda$ is actually GUT-scale. That would start
inflation in the early universe without the need to assume an unnaturally 
smooth initial condition. Inflation would be stopped in this model by the
back-reaction from long wavelength, virtual gravitons which are continually
ripped out of the vacuum by the accelerated expansion of inflation. The
kinetic energy density of these gravitons is exactly diluted by the expansion
of the 3-volume to produce a constant energy density. However, the next 
order effect, due to the gravitational interaction energy between gravitons, 
must slow inflation because gravity is attractive. This next order effect 
grows without bound, so it must eventually stop inflation if nothing else
supervenes first. 

One can understand the growth of back-reaction from the fact that the 
Newtonian potential is $-G M/R$, with the total kinetic energy of infrared 
gravitons growing like $M \sim a^3(t)$ and the radius growing like $R \sim 
a(t)$. Hence the universe must eventually fall inside its own Schwarzschild 
radius! That estimate (of $a^2(t)$ growth for the energy density of
interaction) assumes the newly created gravitons are instantly in contact 
with one another, whereas the self-interaction actually requires time to 
build up. When this causality delay is taken into account the interaction 
energy density grows like $-G H^6 \ln(a)$ at lowest order \cite{TW11}. This 
slower growth means that inflation can only be checked after an {\it 
enormous} number of e-foldings: $\ln(a) \sim 1/(G H^2) \sim 10^{12}$. There 
seems to be no problem with such a large number of e-foldings; indeed, it 
can be viewed as a way of exploiting the weakness of the gravitational 
interaction to make inflation last a long time. We will see that it is also
a way of keeping spatial variation small even over volumes as large as the
currently observable universe.

Although we emphasize that the results of the present work stand on their
own, they can be viewed as a partial check on the ``null hypothesis'' 
that infrared gravitons make no significant corrections to inflationary
quantum gravity \cite{GT,TW3}. Further, if one assumes that the spatially 
homogeneous, leading infrared logarithms stop inflation, then the spatially 
inhomogeneous corrections at the next order are suppressed by one factor of 
$G H^2$, which is the observed strength of the spectrum of primordial 
density perturbations. Hence scalar density perturbations of the right
strength seem to be possible without a fundamental scalar.

Section 3 motivates and defines the Effective Scale Factor Approximation.
In Section 4 we compute the effective scale factor, both perturbatively 
using dimensional regularization at 1- and 2-loop orders and, implicitly, 
at arbitrary order. Section 5 sees this result used to evaluate an invariant 
observable which has been proposed to quantify the quantum gravitational 
back-reaction on inflation \cite{TW4}. Our conclusions comprise Section 6.

Before closing this Introduction we should clarify the distinction between 
``full quantum gravity'' and ``quantum gravity at leading logarithm order''. 
Because general relativity is not perturbatively renormalizable, it cannot 
provide a complete theory of quantum gravity on the perturbative level. 
There must either be some different model, possibly not even based upon a
metric field, or else quantum gravity computations are intrinsically 
nonperturbative. However, failing to determine {\it everything} is not quite 
the same thing as failing to determine {\it anything}. It is perfectly 
valid to employ perturbative quantum general relativity as an effective 
field theory to study phenomena which are driven by infrared gravitons.
Further, the results so obtained cannot be changed by the still unknown, 
ultraviolet completion of the theory. So we will many times derive results 
which take the form,
\begin{equation}
\Bigl({\rm finite\ number}\Bigr) \times \ln(a) + 
\Bigl({\rm divergent\ constant}\Bigr) \; ,
\end{equation}
and retain the infrared logarithm while ignoring the divergent constant. 
This is correct because the infrared logarithm derives exclusively from
long wavelength virtual gravitons, which must be reliably described by 
quantum general relativity. In contrast, the divergent constant originates
in the ultraviolet sector, which cannot be correct, at least not on the
perturbative level. Because no divergences are observed in nature we know 
that the ultraviolet completion of quantum gravity is somehow avoiding them. 
As long as the result is a finite constant, the infrared logarithm must 
eventually dominate at late times. 

This use of a flawed (or misunderstood) formalism as an effective quantum
field theory to infer valid results from the infrared has a long and
distinguished history. The oldest example is the solution of the infrared
problem in quantum electrodynamics by Bloch and Nordsieck \cite{BN}, long
before that theory's renormalizability was suspected. Weinberg \cite{SW0}
was able to achieve a similar resolution for quantum gravity with zero
cosmological constant. The same principle was at work in the Fermi theory
computation of the long range force due to loops of massless neutrinos by
Feinberg and Sucher \cite{FS}. Matter which is not supersymmetric
generates nonrenormalizable corrections to the graviton propagator at one
loop, but this did not prevent the computation of photon, massless neutrino
and massless, conformally coupled scalar loop corrections to the long range
gravitational force \cite{CDH}. More recently, Donoghue \cite{JFD} has 
touched off a minor industry \cite{MV} by applying the principles of low 
energy effective field theory to compute graviton corrections to the long 
range gravitational force. Our analysis exploits the power of low energy 
effective field theory in the same way, differing from the previous examples 
only in the detail that our background geometry is locally de Sitter rather 
than flat.

\section{Quantum Gravity at Leading Log Order}

The purpose of this section is to describe the leading logarithm
approximation for inflationary quantum gravity and to demonstrate
that the theory in this limit is vastly better behaved than full
quantum gravity. We begin by reviewing the free field expansion of 
quantum gravity on a locally de Sitter background. We then take note 
of some key facts about infrared logarithms. Finally, we show that
only the 1-loop renormalization of the cosmological constant can
contribute at leading logarithm order.

\subsection{The Free Field Expansion}

Because we employ dimensional regularization it is necessary to work
in $D$ spacetime dimensions. With a cosmological constant $\Lambda$, the 
gravitational equations of motion are:
\footnote{Hellenic indices take on spacetime values while Latin 
indices take on the $(D-1)$ space values. Our metric tensor has 
spacelike signature and our curvature tensor equals 
$R^{\alpha}_{~ \beta\mu\nu} \equiv 
\Gamma^{\alpha}_{~\; \nu\beta , \mu} + 
\Gamma^{\alpha}_{~\, \mu\rho} \, \Gamma^{\rho}_{~\, \nu\beta} -
(\mu \leftrightarrow \nu)$.
The Hubble parameter is $H \equiv \sqrt{\frac{1}{D-1} \Lambda}$.}
\begin{equation}
R_{\mu\nu} - \frac12 g_{\mu\nu} R + \frac12 (D-2) \, \Lambda g_{\mu\nu} = 0 
\;\; . \label{eom}
\end{equation}
The free field expansion of perturbative quantum gravity has much 
the same structure on a locally de Sitter background as on flat space. 
Consider the case of synchronous gauge for which the invariant element 
takes the form:
\begin{equation}
ds^2 = -dt^2 + a^2(t) \exp\Bigl[\kappa h(t,\vec{x})\Bigr]_{ij} \;
dx^i dx^j 
\qquad , \qquad 
a(t) \equiv e^{H t} 
\;\; . \label{deS}
\end{equation}
Here, the exponential of the graviton field $h_{ij}(t,\vec{x})$ has
its usual meaning:
\begin{equation}
\exp\Bigl[\kappa h\Bigr]_{ij} \equiv 
\; \delta_{ij} + \kappa h_{ij} + 
\frac12 \kappa^2 h_{ik} h_{kj} + \dots 
\end{equation}
where $\kappa^2 \equiv 16 \pi G$ is the loop counting parameter. 

Just as in flat space we can decompose the graviton field into 
irreducible representations of the rotation group:
\begin{equation}
h_{ij} \equiv h^{TT}_{ij} 
+ \partial_i h^T_j + \partial_j h^T_i 
- \frac1{D\!-\!2} \Bigl[\delta_{ij} \!-\! 
(D\!-\!1) \frac{\partial_i \partial_j}{\nabla^2}\Bigr] h^L 
+ \frac1{D\!-\!2} \Bigl[\delta_{ij} \!-\! 
\frac{\partial_i \partial_j}{\nabla^2}\Bigr] h 
\;\; , \label{hdecomp}
\end{equation}
where the usual conditions distinguish the transverse-traceless and
transverse components:
\begin{equation}
\partial_i h^{TT}_{ij} = h^{TT}_{ii} = \partial_i h^T_i = 0
\;\; .
\end{equation}
Like flat space, the linearized equations of motion imply:
\begin{equation}
h^T_i(t,\vec{x}) = h^L(t,\vec{x}) = h(t,\vec{x}) = 0
\;\; , 
\end{equation}
while the linearized solution for the transverse-traceless components 
can be expanded in spatial plane waves:
\begin{equation}
\chi_{ij}(t, {\vec x}) \equiv
\sqrt{2} \int \frac{d^{D-1} k}{(2\pi)^{D-1}} \;
\theta(k \!-\! H) \sum_{\lambda} \left\{ \,
u(t, k) \, e^{i {\vec k} \dot {\vec x}}
\epsilon_{ij}({\vec k}, \lambda) \, 
\alpha({\vec k}, \lambda) + (cc) \, \right\}
\; . \label{linear}
\end{equation}
In (\ref{linear}), $u(t, k)$ are the mode functions, 
$\epsilon_{ij} ({\vec k}, \lambda)$ is the polarization tensor, 
$\alpha({\vec k}, \lambda)$ the annihilation operator, and 
$(cc)$ stands for complex conjugation. The $\frac12 (D\!-\!3)D$ 
creation and annihilation operators and transverse-traceless 
polarization tensors obey the same relations as in flat space 
\cite{Grish,TW5}:
\begin{eqnarray}
\Bigl[\alpha(\vec{k},\lambda) \, , 
\alpha^{\dagger}(\vec{k}',\lambda')\Bigr]
\!& = &\!  
(2\pi)^{D-1} \,
\delta_{\lambda \lambda'} \; 
\delta^{D-1} \! (\vec{k} \!-\! \vec{k}') 
\;\; , \label{cna} \\
\sum_{\lambda} \epsilon_{ij}({\vec k}, \lambda) \;
\epsilon_{k \ell}^{\ast}({\vec k}, \lambda) 
\!& = &\!
\Pi_{i(k}({\widehat k}) \; \Pi_{\ell) j}({\widehat k}) -
\frac{1}{D-2} \, \Pi_{ij}({\widehat k}) \; \Pi_{k \ell}({\widehat k})
\;\; , \qquad \label{pol} \\
\epsilon_{ii}({\vec k}, \lambda) \, = \, 
k_i \, \epsilon_{ij}({\vec k}, \lambda) 
\!\!& = &\!\! 0
\qquad , \qquad
\epsilon_{ij}({\vec k}, \lambda) \; 
\epsilon_{ij}^{\ast}({\vec k}, \lambda') \, = \, 
\delta_{\lambda \lambda'} 
\;\; , \label{pol2}
\end{eqnarray}
where $\Pi_{ij}({\widehat k})$ is the transverse projection operator:
\begin{equation}
\Pi_{ij}({\widehat k}) \, \equiv \, 
\delta_{ij} \, - \, {\widehat k}_i \, {\widehat k}_j
\;\; , \label{pi}
\end{equation}
and where parenthesized indices are symmetrized.

The main difference from flat space is that the mode functions 
$u(t,k)$ go from simple exponentials to Hankel functions:
\footnote{The system is assumed to be released in free Bunch-Davies
vacuum at $t=0$.}
\begin{equation}
u(t,k) \; = \; i \sqrt{\frac{\pi}{4 H}} \; 
a^{-(\frac{D-1}2)} \;
H^{(1)}_{\frac{D-1}2} \Big(\frac{k}{H a}\Bigr) 
\;\; . \label{BD}
\end{equation}
The restriction to $k \equiv \Vert \vec{k} \Vert \geq H$ in (\ref{linear}) 
is made to avoid an infrared singularity in the free propagator \cite{FP}. 
The physical reason for this singularity is that no causal process would 
allow a local observer to prepare coherent Bunch-Davies vacuum over an 
infinite spatial section. Sensible physics can be regained either by 
employing an initial state for which the super-horizon modes are less 
strongly correlated \cite{AV2}, or else by working on a compact spatial 
manifold such as $T^{D-1}$ for which there are initially no super-horizon 
modes \cite{TW6}, and then making the integral approximation to the mode 
sums. In both cases modes with $k < H$ are effectively absent.

The parallel with flat space persists at higher orders. The fundamental
quantity of the free field expansion is the linearized mode sum (\ref{linear}).
All components of $h_{ij}$ can be expressed in terms of $\chi_{ij}$. The 
$g_{00}$ and $g_{0i}$ constraint equations determine $h^T_i$, $h^L$ and 
$h$ as expansions which start at second order:
\begin{equation}
h^T_i \sim h^L \sim h \sim \kappa \chi^2 + \kappa^2 \chi^3 + \dots
\end{equation}
The dynamical $g_{ij}$ equations define a related expansion for $h_{ij}^{TT}$
which starts at first order:
\begin{equation}
h_{ij}^{TT} \sim \chi_{ij} + \kappa \chi^2 + \kappa^2 \chi^3 + \dots
\end{equation}
These expansions will generally involve integrations against retarded 
Green's functions -- which can themselves be expressed as the commutator 
of two free fields:
\begin{equation}
\Bigl[\hbox{}_{ij} G_{k\ell}\Bigr](x;x') \; = \; 
i\theta(t - t') \Bigl[ \chi_{ij}(x) \, , \chi_{k\ell}(x') \Bigr] 
\;\; .
\end{equation}
A nice diagrammatic representation for the free field expansions has been 
given recently by Musso \cite{Musso}.

The expectation value of any operator can be computed using the free
field expansion. One first expands the operator in powers of the
graviton field $h_{ij}$. Each term in this series gives its own 
expansion in powers of the free field $\chi_{ij}$. One then takes
the expectation value of the resulting sums and integrals of products
of free fields, for which the usual reductions of free field correlators 
apply. In particular, the expectation value of $N$ of the $\chi_{ij}$'s
is zero for $N$ odd, while for $N$ even it is the sum of the $(N-1)!!$
distinct permutations of 2-point correlators.

\subsection{Facts about Infrared Logarithms}

We have already seen the crucial role of infrared logarithms for
perturbative quantum gravity on a locally de Sitter background. These
infrared logarithms derive from explicit factors of $\ln(a)$
\footnote{To simplify the notation we have normalized the initial 
value of the scale factor to be $a_{\rm on} = 1$.} 
in free field correlators \cite{old}, and from integrating the 
retarded Green's functions (times these correlators) that arise in 
the free field expansion \cite{TW7,SW}. Based on this understanding 
one can give a simple rule -- valid for any interaction in any 
theory -- for counting the maximum number of infrared logarithms 
that can arise for each extra power of the coupling constant. 
Consider an interaction with $N$ undifferentiated massless, minimally 
coupled scalars or gravitons:
\begin{equation}
\Delta \mathcal{L} \; \sim \;
({\rm coupling\ constant}) \times
(\varphi,h_{ij})^N \times ({\rm anything}) 
\;\; .
\end{equation}
Here, ``anything'' can involve differentiated $\varphi$ or $h_{ij}$'s,
or other fields altogether. Then, for each additional factor of the 
square of the coupling constant there can be at most $N$ additional 
infrared logarithms \cite{PTsW}. A few examples from scalar models 
on a non-dynamical de Sitter background illustrate the rule 
\cite{phi4,SQED,Yukawa}:
\begin{eqnarray}
\Delta \mathcal{L} \; = \; 
-\frac1{4!} \lambda \varphi^4 \; a^D 
& \Longrightarrow &
\lambda \ln^2(a) 
\;\; , \label{phi4} \\
\Delta \mathcal{L} \; = \;
ie \varphi^* A_{\mu} \partial_{\nu} \varphi \;
\eta^{\mu\nu} a^{D-2} 
& \Longrightarrow & 
e^2 \ln(a) 
\;\; , \\
\Delta \mathcal{L} \; = \;
-f \varphi \overline{\psi} \psi \; a^D 
& \Longrightarrow &
f^2 \ln(a) 
\;\; .
\end{eqnarray}

The basic interaction of the bare quantum gravitational Lagrangian has
long been known for de Sitter background \cite{TW8} and it gives
the result \cite{PTsW}:
\begin{equation}
\Delta \mathcal{L} \; \sim \;
\kappa h \partial h \partial h \; a^{D-2}
\quad \Longrightarrow \quad
G H^2 \ln(a) 
\;\; . \label{tree}
\end{equation}
The counting is the same for gravity + fermions \cite{MW1}:
\begin{equation}
\Delta \mathcal{L} \; \sim \;
\kappa h \overline{\psi} \partial \psi 
\; a^{D-1}
\quad \Longrightarrow \quad
G H^2 \ln(a) 
\;\; .
\end{equation}
Hence, the general form of the fermion field strength (\ref{Z2}) is:
\begin{equation}
Z_2(t) = 1 + \sum_{\ell=1}^{\infty} (G H^2)^{\ell} \Biggl\{ 
c_{\ell,0} \Bigl[\ln(a)\Big]^{\ell} + 
c_{\ell,1} \Bigl[\ln(a)\Bigr]^{\ell-1} +
\dots + c_{\ell,\ell-1} \ln(a)\Biggr\} 
\;\; . \label{genform}
\end{equation}
The constants $c_{\ell,k}$ are pure numbers which are assumed to be of 
order one. The term in (\ref{genform}) involving $[G H^2 \ln(a)]^{\ell}$ 
is the {\it leading logarithm} contribution at $\ell$ loop order; the 
other terms are {\it subdominant logarithms}.

Quantum gravitational perturbation theory breaks down when $\ln(a) \sim 
1/{G H^2}$, at which point the leading infrared logarithms at each loop 
order contribute numbers of order one. In contrast, the subleading 
logarithms are all suppressed by at least one factor of the very small 
number $G H^2 \ltwid 10^{-12}$. So it makes sense to retain only the 
leading infrared logarithms:
\begin{equation}
Z_2(t) 
\quad \longrightarrow \quad
1 + \sum_{\ell=1}^{\infty} c_{\ell,0} \Bigl[G H^2
\ln(a)\Bigr]^{\ell} 
\;\; .
\end{equation}
This is known as the {\it leading logarithm approximation}.

It is important to note that, {\it the operator under study can affect 
the form of its leading logarithm expansion.} For example, we saw from 
expression (\ref{phi4}) that one gets at most two infrared logarithms 
for each extra coupling constant from a $\lambda \varphi^4 a^D$ interaction. 
That rule only gives the maximum, and it does not specify the number of 
infrared logarithms in the lowest order result. Explicit computation 
shows that the leading logarithm expansion of the expectation value of 
$2n$ coincident fields is \cite{TW2}:
\footnote{We shall use henceforth $\langle \! \langle \, O \, 
\rangle \! \rangle$ to indicate that the expectation value of an 
arbitrary operator $O$ is taken in the leading logarithm approximation.} 
\begin{eqnarray}
\lefteqn{\langle \! \langle \,
\Omega \vert \, \varphi^{2n}(x) \, \vert \Omega
\, \rangle \! \rangle \; = \; 
(2n \!-\!1)!!  \left[ \frac{H^2 \ln(a)}{4 \pi^2} \right]^n
\Biggl\{ 1 - \frac{n^2 \!+\! n}{2} \, \frac{\lambda}{36 \pi^2} \ln^2(a) } 
\nonumber \\
& & \hspace{2cm} 
+ \, \frac{35 n^4 \!+\! 170 n^3 \!+\! 225 n^2 \!+\! 74 n}{280} \, 
\left[ \frac{\lambda}{36 \pi^2} \ln^2(a) \right]^2 - \dots 
\Biggr\}
\;\; . \qquad
\end{eqnarray}
In this case the bound of $\ln^2(a)$ for each extra $\lambda$ is saturated
for every $n$, but the $\ln(a)$ dependence of the order $\lambda^0$ result
depends upon $n$. In contrast, the expectation value of the kinetic term
takes the form \cite{phi4}:
\begin{equation}
\langle \! \langle \, \Omega \vert \, g^{\mu\nu} \partial_{\mu} \varphi \,
\partial_{\nu} \varphi \, \vert \Omega \, \rangle \! \rangle \; = \; 
-\frac{3 H^4}{8 \pi^2} \Biggl\{ 1 + \# \lambda \ln(a) + 
O\Bigl( \lambda^2 \ln^3(a) \Bigr) \Biggr\} 
\;\; ,
\end{equation}
where $\#$ is a divergent constant.\footnote{One can either view this
kinetic contribution as part of the stress tensor, in which case it 
requires no composite operator renormalization but it also fails to
contribute at leading logarithm order. Such divergent, subleading 
logarithms are automatically removed by renormalization as a 
consequence of the cancellation of overlapping divergences, but they
can leave calculable, finite remainders. (For worked-out examples see 
\cite{phi4}.) Or one can view the coincident kinetic term as a composite
operator in its own right, which of course requires composite operator
renormalization to produce a finite result. For an explicit example of 
how even the leading logarithm contributions to such composite operators 
can be divergent see section 4 of \cite{gW2}.}
Note that the derivatives in the operator under study not only preclude 
there being any infrared logarithms at zeroth order, they also cause loop 
corrections to be one fewer than the maximum given by the rule (\ref{phi4}). 
So instead of the order $\lambda$ correction possessing two infrared 
logarithms, it has only one; instead of the order $\lambda^2$ correction 
having four infrared logarithms, it has only three; and so on.

\subsection{Quantum Gravity}

Working in the leading logarithm approximation effects a dramatic 
simplification in quantum gravity. Full quantum gravity is haunted
by the yet unknown -- and experimentally untested -- nature of its 
ultraviolet completion. In perturbation theory this shows up as an 
escalating series of divergences and counterterms whose arbitrary 
finite parts degrade -- but don't entirely eliminate \cite{JFD}
-- the theory's ability to make predictions. In contrast, the leading 
logarithm terms of quantum gravity are largely dominated by the 
infrared sector, whose preservation is an essential correspondence 
limit of any ultraviolet completion. And the only counterterm that 
contributes at leading logarithm order is the 1-loop renormalization 
of the cosmological constant.

This crucial insight concerning counterterms derives from the rule
given above. The cosmological counterterm is:
\begin{equation}
\Delta \mathcal{L}_1 \; \equiv \;
-\frac{(D\!-\!2) \delta \Lambda}{16 \pi G} \, \sqrt{-g} 
\;\; .
\end{equation}
We can express $\delta \Lambda$ as dimensionless contributions from one 
loop ($\delta \Lambda_1$), two loops ($\delta \Lambda_2$) and so forth:
\begin{equation}
\delta \Lambda \; \equiv \;
H^2 \Bigl\{ \, \delta \Lambda_1 \times \kappa^2 H^2 \, + \,
\delta \Lambda_2 \times (\kappa^2 H^2)^2 \, + \, 
\dots \, \Bigr\} 
\;\; .
\end{equation}
The measure factor has its standard expansion:
\begin{equation}
\sqrt{-g} \; = \;
a^D \Bigl\{ \, 1 + \frac12 \kappa h_{ii} + 
\frac18 \kappa^2 (h_{ii})^2 + \dots \, \Bigr\} 
\;\; .
\end{equation}
It follows that the cosmological counterterm has the general form:
\begin{equation}
\Delta \mathcal{L}_1 \; \sim \;
H^4 \Bigl\{ \, \delta \Lambda_1 \, + \,
\delta \Lambda_2 \times \kappa^2 H^2 \, + \, \dots \, \Bigr\} 
\times \kappa^n h^n a^D 
\;\; .
\end{equation}
The 1-loop term has the same number of undifferentiated gravitons
per coupling constant as the bare interaction (\ref{tree}), so it 
contributes at leading logarithm order. However, one can see that 
the 2-loop term -- the one proportional to $\delta \Lambda_2$ -- is 
suppressed by a factor of the minuscule parameter $\kappa^2 H^2$. It 
contributes to the first subleading logarithm term, and is completely 
absent from the leading logarithm approximation. The higher loop 
contributions to $\delta \Lambda$ are even more suppressed.

The higher counterterms of quantum gravity can be organized so that 
their graviton expansions always contain differentiated fields:
\begin{eqnarray}
\Delta \mathcal{L}_2 
& \equiv & 
-\frac{\delta G}{16 \pi G^2} \, 
\Bigl[ R - (D\!-\!2) \Lambda \Bigr] \sqrt{-g} 
\;\; , \\
\Delta \mathcal{L}_3 
& \equiv & 
\alpha \, \Bigl[ R - D \Lambda \Bigr]^2 \sqrt{-g} 
\;\; , \\
\Delta \mathcal{L}_4 
& \equiv & 
\beta \, C^{\rho\sigma\mu\nu} C_{\rho\sigma\mu\nu} \sqrt{-g} 
\;\; ,
\end{eqnarray}
and so on. As with $\delta \Lambda$, we can express $\delta G$ in 
terms of dimensionless, $\ell$-loop contributions $\delta G_{\ell}$:
\begin{equation}
\delta G \; \equiv \;
\kappa^2 \Bigl\{ \, \delta G_1 \times \kappa^2 H^2 \, + \,
\delta G_2 \times (\kappa^2 H^2)^2 \, + \, \dots \, \Bigr\} 
\;\; .
\end{equation}
Hence, the Newtonian constant counterterm takes the form:
\begin{equation}
\Delta \mathcal{L}_2 \; \sim \; 
\kappa^2 H^2 \Bigl\{ \, \delta G_1 + 
\delta G_2 \times \kappa^2 H^2 + \dots \,\Bigr\} 
\Bigl\{ \, \partial h \partial h + 
\kappa h \partial h \partial h + \dots \, \Bigr\} \, 
a^{D-2} 
\;\; .
\end{equation}
When compared with the bare interaction (\ref{tree}) even the 
1-loop contribution has an extra factor of $\kappa^2 H^2$ for each 
undifferentiated graviton. Therefore, the Newtonian constant 
counterterm makes no contribution at leading logarithm order, 
and the same holds for {\it all} higher counterterms.

The reason for this amelioration of the ultraviolet problem is that 
undifferentiated fields effectively lose their ultraviolet modes at 
leading logarithm order. One can only reach leading logarithm order 
if every term in the free field expansion that can contribute a 
factor of $\ln(a)$ actually does so. These factors of $\ln(a)$ come 
only from the infrared sector, that is, from $H < k < H a(t)$ in the 
free field mode sum (\ref{linear}) \cite{PTsW}. For example, the 
expectation value of two coincident free fields is:
\begin{eqnarray}
\lefteqn{ \Bigl\langle \, \Omega \Bigl\vert \, 
\chi_{ij}(t,\vec{x}) \, \chi_{k\ell}(t,\vec{x}) 
\, \Bigr\vert \Omega \, \Bigr\rangle
\; = \; \frac{2 D (D\!-\!3)}{(D\!+\!1) (D\!-\!2)} 
\left[ \delta_{i (k} \delta_{\ell) j} - 
\frac1{D\!-\!1} \, \delta_{ij} \delta_{k\ell} \right] }
\nonumber \\
& & \hspace{4cm} 
\times \, \frac1{2^{D-2} \, \pi^{\frac{D-1}2} \, \Gamma(\frac{D-1}2)}
\int_H^{\infty} dk \, k^{D-2} \, \Bigl\Vert u(t,k) \Bigr\Vert^2 
\;\; , \qquad \\
& & = \frac{2 D (D\!-\!3)}{(D\!+\!1) (D\!-\!2)} 
\left[ \delta_{i (k} \delta_{\ell) j} - 
\frac1{D\!-\!1} \, \delta_{ij} \delta_{k\ell} \right] 
\nonumber \\
& & \hspace{4cm} 
\times \, \frac{H^{D-2}}{(4\pi)^{\frac{D}2}} \,
\left\{ (UV) \, + \, 
\frac{\Gamma(D\!-\!1)}{\Gamma(\frac{D}2)} \, 
2 \ln(a) \right\} 
\;\; . \qquad \label{keylog}
\end{eqnarray}
To get the ultraviolet divergent constant $(UV)$ requires the full
range of integration and the complete form of the integrand. However, 
this divergent constant is guaranteed to be of subleading logarithm order,
so it holds no interest for us. To extract the finite infrared logarithm 
it suffices to retain only the first term in the long wavelength expansion 
of the mode function:
\begin{equation}
k^{D-2} \Bigl\Vert u(t,k) \Bigr\Vert^2 \; = \;
\frac{\Gamma^2(\frac{D-1}2) \, 2^{D-3} \, H^{D-2}}{\pi k} 
\left\{1 \, + \, O\Bigl(\frac{k^2}{H^2 a^2}\Bigr) \right\} 
\;\; .
\end{equation}
We can furthermore restrict the range of integration to just the infrared:
\begin{eqnarray}
\frac1{2^{D-2} \, \pi^{\frac{D-1}2} \, \Gamma(\frac{D-1}2)} 
\int_H^{\infty} dk \, k^{D-2} \, \Bigl\Vert u(t,k) \Bigr\Vert^2 
& \longrightarrow &
\frac{\Gamma(\frac{D-1}2) H^{D-2}}{2 \pi^{\frac{D+1}2}} 
\int_{H}^{H a} \frac{dk}{k} 
\; , \qquad \\ 
& = & \!\!\!
\frac{H^{D-2}}{(4 \pi)^{\frac{D}2}} \, 
\frac{\Gamma(D\!-\!1)}{\Gamma(\frac{D}2)} \; 2 \ln(a) 
\; .
\end{eqnarray}

The simplifications noted above do not apply to the correlators of
differentiated free fields. Because these correlators cannot contribute
infrared logarithms, the constants they contribute (times one power
of $a$ for every spatial derivative) come from all parts of the mode 
sum and from all terms in the mode function. It follows that {\it the 
correlators of differentiated free fields must be evaluated exactly, 
with the dimensional regularization in place.}

It is well to close this section with a summary of facts about infrared
logarithms in quantum gravity:
\begin{itemize}
\item{There can be at most one factor of $\ln(a)$ for each extra factor
of the loop counting parameter $G H^2$;}
\item{The operator under study controls how many infrared logarithms
occur at lowest order, and it may also cause the number expected at
higher orders to be fewer than the maximum just noted;}
\item{The only counterterm that affects the leading logarithm order is 
the 1-loop renormalization of the cosmological constant; and}
\item{Even at leading logarithm order the correlators of differentiated 
free fields must be computed exactly, with the (dimensional) regularization
in place.}
\end{itemize}

\section{Effective Scale Factor Approximation}

The purpose of this section is to explain the Effective Scale Factor
Approximation. We begin by simply defining the approximation. We then 
give a motivation for it. Finally, we comment on the potential problem 
of using expectation values rather than stochastic samples, and the 
closely related issue of spatial inhomogeneities.

\subsection{The Approximation}

The dynamical variable of synchronous gauge quantum gravity is the 
spatial metric, $g_{ij}(t,\vec{x})$. On de Sitter background it is 
natural to express this as follows in terms of the graviton field 
$h_{ij}(t,\vec{x})$:
\begin{equation}
g_{ij}(t,\vec{x}) \; \equiv \;
a^2(t) \exp\Bigl[\kappa h(t,\vec{x})\Bigr]_{ij} 
\qquad , \qquad 
a(t) = e^{H t} 
\;\; .
\end{equation}
We have seen that the various components of the graviton field can be 
expanded in powers of the transverse-traceless free field 
$\chi_{ij}(t,\vec{x})$, which is defined by expressions 
(\ref{linear}-\ref{BD}). We also saw from (\ref{keylog}) that infrared 
logarithms derive from the steady growth in the amplitude of $\chi_{ij}$. 

The complete expansion of $h_{ij}$ in powers of $\chi_{ij}$ reflects 
the full complexity of perturbative quantum gravity, including the 
ultraviolet. It is neither possible to retain all this complexity, nor
would it even be desirable in the absence of fully understanding the 
physical ultraviolet completion of the theory. What we seek instead is 
a simplified expansion that reproduces the leading infrared logarithms. 
We propose that the {\it Effective Scale Factor Approximation} may 
provide such an expansion. The basic idea is that the graviton remains
transverse-traceless and free, but propagates in the background
geometry of an effective scale factor $A(t)$ which is determined from
the expectation value of the $g_{00}$ gravitational constraint equation
of motion. More precisely, this approximation is defined by three 
statements:
\begin{enumerate}
\item{The full metric is the square of a $\comp$-number, effective
scale factor $A(t)$ times the exponential of a transverse-traceless
mode sum $H_{ij}[A](t,\vec{x})$:
\begin{equation}
g_{ij}(t,\vec{x}) 
\quad \longrightarrow \quad
A^2(t) \, \exp \left[ \kappa H[A](t,\vec{x}) \right]_{ij} 
\;\; . \label{metric}
\end{equation}}
\item{The transverse-traceless mode sum $H_{ij}[A](t,\vec{x})$ is the 
same as the free field $\chi_{ij}(t,\vec{x})$ with the de Sitter mode
functions (\ref{BD}) replaced by the functions $u[A](t,k)$ that pertain 
for scale factor $A(t)$ \cite{TW9}:
\begin{eqnarray}
\lefteqn{H_{ij}[A](t,\vec{x}) \; \equiv } 
\nonumber \\
& \mbox{} & \hspace{-1cm}
\sqrt{2} \int \frac{d^{D-1}k}{(2\pi)^{D-1}} \;
\theta(k \!-\! H) \sum_{\lambda} \left\{ \,
u[A](t,k) \, e^{i {\vec k} \dot {\vec x}}
\epsilon_{ij}({\vec k}, \lambda) \, 
\alpha({\vec k}, \lambda) + (cc) \, \right\}
\; . \qquad \label{newgrav}
\end{eqnarray}}
\item{The effective scale factor $A(t)$ is determined from the leading
logarithm expectation value of the $g_{00}$ equation of motion 
(\ref{eom}), including the 1-loop renormalization of the cosmological 
constant:
\begin{equation}
\Bigl\langle \!\! \Bigl\langle \, \Omega \Bigl\vert \, 
R_{00} + \frac12 R - \frac12 (D\!-\!2) 
\Bigl( \Lambda \!+\! \delta \Lambda_1 \kappa^2 H^4 \Bigr) 
\, \Bigr\vert \Omega \, \Bigr\rangle \!\! \Bigr\rangle
\; = \; 0 
\;\; . \label{Aeqn}
\end{equation}
Note that the expectation value of any operator in the presence of a 
homogeneous and isotropic state such as the vacuum must itself be
homogeneous and isotropic, even if the operator is not, so (\ref{Aeqn})
can indeed be regarded as an equation for $A(t)$.}
\end{enumerate}

\subsection{Motivation}

In terms of the graviton's general decomposition (\ref{hdecomp}) it will
be seen that the Effective Scale Factor Approximation amounts to the 
following assumptions about the behavior at leading log order:
\begin{eqnarray}
h_{ij}^{TT}(t,\vec{x}) 
& \longrightarrow & 
H_{ij}[A](t,\vec{x}) 
\;\; , 
\label{hTT} \\
h^T_i(t,\vec{x}) 
& \longrightarrow & 
0 
\;\; , \label{hT} \\
h^L(t,\vec{x}) 
& \longrightarrow & 
\ln \! \left[ \frac{A^2(t)}{a^2(t)} \right] 
\;\; , \label{hL} \\ 	
h(t,\vec{x}) 
& \longrightarrow & 
(D\!-\!1) \, \ln \! \left[ \frac{A^2(t)}{a^2(t)} \right] 
\;\; . \label{htrace}
\end{eqnarray}
Because we do not yet know how to derive these assumptions from the
leading logarithm approximation it is best to regard them, at this
stage, as being {\it in addition} to it. However, we susepct that a
proper derivation exists and we advance the following reasons for 
this belief:
\begin{itemize}
\item{A key distinction between gravity and the various scalar models 
\cite{phi4,SQED,Yukawa} that have been studied is that a constant value 
of the graviton field has no gauge invariant significance. Unlike
scalar theories, gravitation cannot develop an effective potential other 
than the cosmological constant itself. There are certainly quantum
corrections, but they must always involve derivatives, which reduces the
number of infrared logrithms. This suggests that the leading logarithm
theory may be ``free'' in some variable. A plausible candidate for this 
variable is $h^{TT}_{ij}(t,\vec{x})$ because it is the only component of 
the graviton that involves $\chi_{ij}(t,\vec{x})$ at first order.}
\item{By rotational covariance the transverse field $h^T_i(t,\vec{x})$ 
must involve a spatial derivative, which should preclude this component 
of the graviton from ever contributing at leading logarithm order.}
\item{Note that constraint equations determine $h^L(t,\vec{x})$ and 
$h(t,\vec{x})$ as functionals of $h^{TT}_{ij}$. The full solutions
are certainly not spatially homogeneous. However, every spatial 
inhomogeneity means that a pair of transverse-traceless free fields has 
failed to contribute an infrared logarithm, which is spatially homogeneous. 
Hence the leading logarithm result should take the form 
(\ref{hL}-\ref{htrace}) we have assumed.}
\item{The $g_{00}$ constraint equation of motion would ordinarily 
determine a linear combination of $h^L(t,\vec{x})$ and $h(t,\vec{x})$, 
so it is natural to use it to infer the effective scale factor $A(t)$. 
Taking the expectation value at leading logarithm order ensures 
consistency of the method.}
\end{itemize}

It will be seen that the Effective Scale Factor Approximation provides
the possibility for a self-consistent test of the ``null hypothesis''
that infrared logarithms make no significant corrections to inflationary
quantum gravity \cite{GT}. For if the null hypothesis were correct then
either the effective scale factor should remain at its de Sitter value ---
and the graviton should remain free --- or else there are corrections but 
they drop out of a gauge invariant measure of back-reaction. In sections 
4 and 5 we show that neither supposition is correct.

Even though the long wavelength limit of a graviton is gauge equivalent 
to zero, it will be seen that {\it each term in the graviton expansion 
of equation (\ref{Aeqn}) contains two differentiated gravitons}. So any 
effect is sourced by the nonzero values of these derivatives and would 
indeed vanish for a purely constant graviton. However, recall that the 
expectation values of differentiated free fields cannot be infrared 
truncated, so one has instead:
\begin{eqnarray}
\Bigl\langle \, \Omega \Bigl\vert \, 
\dot{\chi}_{ij}(t,\vec{x}) \, \dot{\chi}_{k\ell}(t,\vec{x}) 
\, \Bigr\vert \Omega \, \Bigr\rangle 
& \sim &
H^4 
\;\; , \\
\Bigl\langle \, \Omega \Bigl\vert \,
\partial_m \chi_{ij}(t,\vec{x}) \, 
\partial_n \chi_{k\ell}(t,\vec{x}) 
\, \Bigr\vert \Omega \, \Bigr\rangle 
& \sim &
a^2(t) H^4 
\;\; .
\end{eqnarray}
These are still only small corrections, but they get enhanced by infrared
logarithms at higher orders and the final effect may become large at late 
times. 

\subsection{VEVs versus Stochastic Samples}

Inflationary cosmologists sometimes object to taking expectation
values to follow evolution of a homogeneous, mean geometry, as we
do in (\ref{Aeqn}). They maintain that a long period of cosmological
evolution presents a sort of Schrodinger Cat Paradox in which the 
wave function of the universe decoheres into many different portions, 
which are no longer in good quantum contact with one another, and 
in which the geometry is not even approximately spatially homogeneous 
on super-horizon scales \cite{Linde}. There is no question that this 
view is correct at some level because the vacuum expectation value of 
the stress-energy tensor is homogeneous and isotropic (because the
state is) whereas we perceive (indeed we {\it are}) inhomogeneities
and anisotropies. The key issues which have not been quantitatively 
addressed, and which bear on the validity of the present analysis,
are:
\begin{enumerate}
\item{How unreliable are expectation values? and}
\item{How much spatial variation should one expect?}
\end{enumerate}

We suspect that the reliability of expectation values depends upon 
the operator under study. For operators which average to zero, such
as the density perturbation, the entire result arises from the 
decoherence effect, so one makes an enormous mistake in ignoring it. 
Other operators --- for example, the square of a scalar field --- 
acquire a significant homogeneous expectation value upon which spatial 
variations are superimposed. Any quantum fluctuation drives this 
sort of operator positive, so one might happen to inhabit a special 
region of the universe in which there is little effect for a long 
time, but there will sooner or later be a large effect. The 
expectation value of such an operator can correctly reflect the 
long-term trend everywhere in space, even though it misses variations 
from one region to another.

The example of $\varphi^2(t,\vec{x})$ is not specious because the
free expectation value of this operator is one of the two ways 
infrared logarithms arise in perturbative computations. An even 
more likely candidate for the general reliabilty of expectation
values is the gravitational interaction between the infrared 
gravitons which everyone agrees are produced during inflation.
The self-gravitation of these particles must slow inflation, even
if their distribution is not perfectly homogeneous. And this 
self-gravitation must grow with time as more and more infrared
gravitons are ripped out of the vacuum by inflation. As discussed 
in section 1, the fact that the total kinetic energy of infrared 
gravitons grows like the cube of the universe's radius means that
the entire universe rather quickly falls inside its own 
Schwarzschild radius. The actual gravitational potential grows like 
$\ln(a)$, rather than $a^2$, because causality delays effects from a 
patch of spacetime until it passes within one's past light-cone
\cite{TW11}. Because gravity is a weak interaction, even at the
scales of primordial inflation, the self-gravitation of inflationary
gravitons cannot be significant until the past light-cone has become
enormous. So a significant back-reaction from quantum gravity can
only occur as the cumulative effect of very many small fluctuations
within the past light-cone of the observation point. Because the past 
light-cones of nearby points must largely overlap, there cannot be 
much spatial variation within a few Hubble radii. There might well be
significant variation on very large scales, but even this should not 
change the long-term slowing trend everywhere. So we believe that 
quantifying this trend through the use of expectation values does not 
represent a serious error.

We emphasize that the reliability of expectation values is a quantitative
issue rather than a qualitative one. We do {\it not} maintain that the
actual universe is exactly homogeneous and isotropic, as its quantum
average is. Rather we argue that using expectation values is unlikely to
produce significant errors for the specific back-reaction effect we seek 
to study. Far from ignoring it, the inherently stochastic spatial variation 
of back-reaction plays an essential role in our model of inflation: {\it it 
is how one can get scalar density perturbations of about the correct 
strength without the need for a fundamental scalar.}

To understand what we have in mind, suppose that the leading infrared
logarithms grow to become nonperturbatively strong and that they sum up to
produce a graceful exit to inflation. As noted above, there may be large 
spatial variations on vastly super-horizon scales, but the general trend 
should be correctly described by the average geometry. Because the slowing
effect derives from an enormous number of small fluctuations generated over 
the past light-cone of a very large ($\sim 10^{12}$) number of e-foldings, 
there can be only small spatial variation, even over regions as large as 
the currently observable universe. (Because the $\sim 60$ e-foldings after 
the end of primordial inflation is very much smaller than the $\sim 10^{12}$
we envisage to end inflation.) But there is {\it some} spatial variation; 
what is its likely strength? Because the leading infrared logarithms are 
spatially homogeneous and isotropic, perturbations around them must be 
subdominant. In other words, inhomogeneities must be suppressed by at least 
one power of $G H^2$. Because the leading logarithms are assumed to give an
order one effect (stopping inflation) the first subleading terms should 
induce a spectrum of density perturbations of order $G H^2$. {\it This is 
what one wants!} The power spectra for scalar and tensor perturbations
of wave number $k$ can be approximately expressed in terms of the values 
of the Hubble parameter and deceleration parameter at the time $t_k$
when that mode experienced first horizon crossing \cite{TW9}
\begin{equation}
\mathcal{P}_{\rm S}(k) \sim \frac{G H^2(t_k)}{1 + q(t_k)} \qquad , \qquad
\mathcal{P}_{\rm T}(k) \sim G H^2(t_k) \; .
\end{equation}
There is no way to recognize a factor such as $1/[1 + q(t_k)]$, and this
level of analysis, but the factors of $G H^2(t_k)$ come out exactly right.

\section{Computing the Effective Scale Factor}

The purpose of this section is to solve equation (\ref{Aeqn}) for the
effective scale factor $A(t)$. Because differentiated fields must be
treated differently than undifferentiated ones, we begin by isolating 
the derivatives. We then explain how the various indices can be factored
out to reduce any 2-point correlator to one involving two massless,
minimally coupled scalars. Next we give a brief description of how the 
equation can be formulated for a nonperturbative, numerical evolution. 
We close by working out the 1- and 2-loop corrections.

\subsection{Isolating the Derivatives}

In synchronous gauge ($g_{00} = -1$, $g_{0i} = 0$) the purely temporal
component of the Einstein tensor $G_{\mu\nu}$ takes the form:
\begin{equation}
G_{00} \; \equiv \;
R_{00} - \frac12 g_{00} R \; = \;
\frac18 \, g^{ij} g^{k\ell} ( \, 
\dot{g}_{ij} \dot{g}_{k \ell} - \dot{g}_{ik} \dot{g}_{j\ell} \, ) 
+ \frac12 \, \mbox{}^{(D-1)\!}R 
\;\; ,
\end{equation}
where $\mbox{}^{(D-1)\!}R$ is the Ricci scalar constructed from the 
$(D\!-\!1)$-dimensional spatial metric $g_{ij}(t,\vec{x})$. We now
extract the effective scale factor:
\begin{equation}
g_{ij}(t,\vec{x}) \, \equiv \, 
A^2(t) \, {\bar g}_{ij}(t,\vec{x}) 
\;\; , \label{spaceg}
\end{equation}
so that:
\begin{eqnarray}
\lefteqn{G_{00} \; = \;
\frac12 (D\!-\!1)(D\!-\!2) \frac{\dot{A}^2}{A^2} \, + \, 
\frac12 (D\!-\!2) \frac{\dot{A}}{A} \, {\bar g}^{ij} \dot{\bar g}_{ij} }
\nonumber \\
& & \hspace{2.5cm} + \, \frac18 \, {\bar g}^{ij} {\bar g}^{k\ell} 
( \, \dot{\bar g}_{ij} \dot{\bar g}_{k \ell} - 
\dot{\bar g}_{ik} \dot{\bar g}_{j\ell} \, ) \, + \, 
\frac1{2 A^2} \, \mbox{}^{(D-1)\!}{\bar R} 
\;\; ,
\end{eqnarray}
where the spatial Ricci scalar is:
\begin{eqnarray}
\mbox{}^{(D-1)\!}{\bar R} 
& \equiv &
{\bar g}^{ij} {\bar g}^{k\ell} \Bigl[ \, 
{\bar \Gamma}_{\ell ji,k} - 
{\bar \Gamma}_{\ell ki,j} \, \Bigr] + 
{\bar g}^{ij} {\bar g}^{k\ell} {\bar g}^{mn} 
\Bigl[ \, {\bar \Gamma}_{mjk} {\bar \Gamma}_{n\ell i} - 
{\bar \Gamma}_{mij} {\bar \Gamma}_{nk\ell} \, \Bigr] 
\;\; , \\
& = &
{\bar g}^{ij} {\bar g}^{k\ell} \Bigl[ \,
{\bar g}_{ik,j\ell} - {\bar g}_{ij,k\ell} \, \Bigr] 
\nonumber \\ 
& \mbox{} & \hspace{-1.5cm}
+ \, {\bar g}^{ij} {\bar g}^{k\ell} {\bar g}^{mn} 
\left[ \, \frac14 \, {\bar g}_{ik,m} \, {\bar g}_{j\ell,n} - 
\frac14 \, {\bar g}_{ij,m} \, {\bar g}_{k\ell,n} - 
{\bar g}_{ik,\ell} \, {\bar g}_{jm,n} + 
{\bar g}_{ik,\ell} \, {\bar g}_{mn,i} \right] 
\;\; . \qquad
\end{eqnarray}

At this point we are ready to substitute the Effective Scale Factor
Approximation form (\ref{metric}) for the metric. The fact that 
${\bar g}_{ij}$ is the exponential of a traceless field means that: \\
{\it (i)} we can drop all derivatives of the form ${\bar g}^{ij} 
\partial {\bar g}_{ij}$, \\
{\it (ii)} we can convert the second derivative terms to a total 
derivative -- whose expectation value must vanish -- plus products 
of first derivatives:
\begin{eqnarray}
{\bar g}^{ij} {\bar g}^{k\ell} \Bigl[ \, 
{\bar g}_{ik,j\ell} - {\bar g}_{ij,k\ell} \, \Bigr] 
& = &
\nonumber \\
& \mbox{} & \hspace{-2.5cm}
- \, {\bar g}^{ij}_{~~ , ij} + \,
{\bar g}^{ij} {\bar g}^{k\ell} {\bar g}^{mn} \Bigl[ \,
{\bar g}_{i k , \ell} \, {\bar g}_{j m ,n} + 
{\bar g}_{i k , m} \, {\bar g}_{n \ell , j} - 
{\bar g}_{i k , m} \, {\bar g}_{ j \ell , n} \, \Bigr] 
\;\; . \qquad
\end{eqnarray}
As a result, the equation for $A(t)$ becomes:
\begin{eqnarray}
0 & = & 
\Bigl\langle \!\! \Bigl\langle \,
\Omega \Bigl\vert \, G_{00} \, \Bigr\vert \Omega 
\Bigr\rangle \!\! \Bigr\rangle -
\frac12 (D\!-\!2) \Bigl(\, \Lambda + \delta \Lambda_1 
\kappa^2 H^4 \, \Bigr) 
\;\; , \\
& = & 
\frac12 (D\!-\!2) \left[ 
(D\!-\!1) \frac{\dot{A}^2}{A^2} - \Lambda -  
\delta \Lambda_1 \kappa^2 H^4 \right] - 
\frac18 \Bigl\langle \!\! \Bigl\langle \, 
\Omega \Bigl\vert \, {\bar g}^{ij} {\bar g}^{k\ell} \,
\dot{\bar g}_{jk} \, \dot{\bar g}_{\ell i} 
\, \Bigr\vert \Omega \Bigr\rangle \!\! \Bigr\rangle
\qquad \nonumber \\
& \mbox{} &
+ \, \frac1{2 A^2} \, \Bigl\langle \!\! \Bigl\langle
\Omega \Bigl\vert \, 
{\bar g}^{ij} {\bar g}^{k\ell} {\bar g}^{mn} 
\Bigl( \, - \frac34 {\bar g}_{ik,m} \, {\bar g}_{j\ell,n} + 
{\bar g}_{i k,m} \, {\bar g}_{n \ell, j} \, \Bigr) 
\, \Bigr\vert \Omega \Bigr\rangle \!\! \Bigr\rangle
\;\; . \label{inteqn}
\end{eqnarray}

It remains to act the derivatives on specific $H_{ij}[A](t,\vec{x})$ 
fields and isolate the expectation values of the differentiated fields. 
For this purpose, it is useful to recall a standard formula for the 
derivative of the exponential of a matrix $M_{ij}$:
\begin{equation}
\partial e^{M} = \, e^{M} \Biggl\{ 
\partial M + \frac1{2!} [\partial M, M] + 
\frac1{3!} \Bigl[ [\partial M , M], M\Bigr] + \dots \Biggr\}
\;\; .
\end{equation}
We can define the term on the right hand side as the differentiated 
matrix contracted into a 4-index object $B[M]_{ijk\ell}$:
\begin{equation}
\partial \Bigl( e^{M}\Bigr)_{ij} \; \equiv \;
\Bigl( e^{M}\Bigr)_{ik} \times B[M]_{kjpq} \times \partial M_{pq} 
\;\; , 
\end{equation}
which implies:
\begin{eqnarray}
\lefteqn{B[M]_{ijk\ell} \; = \; 
\delta_{ik} \delta_{\ell j} + 
\frac1{2!} \Bigl[ \, 
\delta_{ik} M_{\ell j} - M_{ik} \delta_{\ell j} 
\, \Bigr] } 
\nonumber \\
& & \hspace{3.5cm} 
+ \, \frac1{3!} \Bigl[ \, 
\delta_{i k} (M^2)_{\ell j} - 2 M_{ik} M_{\ell j} + 
(M^2)_{ik} \delta_{\ell j} \, \Bigr] + 
\dots \qquad \label{B}
\end{eqnarray}

The 4-index symbol $B[\kappa H]_{ijk\ell}$ is particularly useful for
expressing derivatives of ${\bar g}_{ij} = \exp[\kappa H]_{ij}$ because
the differentiated metrics in (\ref{inteqn}) are contracted into 
inverse metrics:
\begin{eqnarray}
{\bar g}^{ij} {\bar g}^{k\ell} 
\dot{\bar g}_{jk} \, \dot{\bar g}_{\ell i} 
& = & 
B[\kappa H]_{ijk\ell} \, B[\kappa H]_{jipq} \times
\kappa^2 \dot{H}_{k\ell} \, \dot{H}_{pq} 
\;\; , \label{dtrel} \\
{\bar g}^{ij} {\bar g}^{k\ell} {\bar g}^{mn} 
{\bar g}_{ik,m} \, {\bar g}_{j\ell , n} 
& = & 
B[\kappa H]_{ijk\ell} \, B[\kappa H]_{jipq} \, 
{\bar g}^{mn} \times 
\kappa^2 H_{k\ell , m} \, H_{pq , n} 
\;\; , \\
{\bar g}^{ij} {\bar g}^{k\ell} {\bar g}^{mn} 
{\bar g}_{ik,m} \, {\bar g}_{n \ell , j} 
& = & 
B[\kappa H]_{nik\ell} \, B[\kappa H]_{mjpq} \,
{\bar g}^{ij} \times 
\kappa^2 H_{k\ell , m} \, H_{pq , n} 
\;\; . \qquad
\end{eqnarray}
It follows that the equation for the effective scale factor can 
be written as:
\begin{eqnarray}
\frac12 (D\!-\!2) \left[ \,
(D\!-\!1) \frac{\dot{A}^2}{A^2} - \Lambda -  
\delta \Lambda_1 \kappa^2 H^4 \, \right]  
& = &
\nonumber \\
& \mbox{} & \hspace{-7cm} 
\frac18 \Bigl\langle \!\! \Bigl\langle
\Omega \Bigl\vert \, 
B[\kappa H]_{ijk\ell} \, B[\kappa H]_{jipq} 
\, \Bigr\vert \Omega \Bigr\rangle \!\! \Bigr\rangle 
\times \Bigl\langle \, \Omega \Bigl\vert \,
\kappa^2 \dot{H}_{k\ell} \, \dot{H}_{pq} 
\, \Bigr\vert \Omega \, \Bigr\rangle 
\nonumber \\
& \mbox{} & \hspace{-7cm} 
+ \, \Bigl\langle \!\! \Bigl\langle
\, \Omega \Bigl\vert \, 
\frac38 \, B[\kappa H]_{ijk\ell} \, B[\kappa H]_{jipq} \, {\bar g}^{mn} 
- \frac12 B[\kappa H]_{nik\ell} \, B[\kappa H]_{mjpq} \, {\bar g}^{ij} 
\, \Bigr\vert \Omega \Bigr\rangle \!\! \Bigr\rangle
\qquad \nonumber \\
& \mbox{} & \hspace{-1cm}  
\times \frac1{A^2} \,
\Bigl\langle \, \Omega \Bigl\vert \,
\kappa^2 H_{k\ell , m} \, H_{pq , n} 
\, \Bigr\vert \Omega \, \Bigr\rangle 
\;\; . \qquad \label{nperteqn}
\end{eqnarray}

\subsection{Factoring Out the Indices}

Since the mode functions $u[A](t,k)$ depend upon $\vec{k}$ only through 
its magnitude, we can use rotational invariance to reduce expectation 
values of $H_{ij}[A](t,\vec{x})$ to those of a fictitious scalar mode sum:
\begin{equation}
\varphi(t,\vec{x}) \equiv 
\int \frac{d^{D-1}k}{(2\pi)^{D-1}} \;
\theta(k \!-\! H) \left\{ \, 
u[A](t,k) \, e^{i\vec{k} \cdot \vec{k}} \, \alpha(\vec{k}) + 
(cc) \, \right\} 
\;\; . \label{scalar}
\end{equation}
This is achieved by writing the expectation value as an integral of the
form:
\begin{equation}
\int \frac{d^{D-1}k}{(2\pi)^{D-1}} \;
f(k) \; {\widehat k}_i \, {\widehat k}_j \cdots 
\;\; ,
\end{equation}
and then extracting the indices through the familiar replacements:
\begin{eqnarray}
{\widehat k}_i \, {\widehat k}_j
& \longmapsto &
\frac{1}{D-1} \, \delta_{ij} 
\;\; , \\
{\widehat k}_i \, {\widehat k}_j \, {\widehat k}_k \, {\widehat k}_{\ell}
& \longmapsto &
\frac{1}{(D+1)(D-1)} \,
\Bigl[ \, \delta_{ij} \, \delta_{k\ell} + 
\delta_{ik} \, \delta_{j\ell} +
\delta_{i\ell} \, \delta_{jk} \, \Bigr] 
\;\; , \\
{\widehat k}_i \, {\widehat k}_j \, {\widehat k}_k \, 
{\widehat k}_{\ell} \, {\widehat k}_m \, {\widehat k}_n
& \longmapsto &
\frac{1}{(D+3)(D+1)(D-1)} \,
\Bigl[ \, \delta_{ij} \, \delta_{k\ell} \, \delta_{mn} +
\cdots \, \Bigr] 
\;\; . \qquad
\end{eqnarray}

>From the mode sum (\ref{newgrav}) and definitions (\ref{cna}-\ref{pol}) 
we can evaluate the coincident 2-point functions of interest: \\
{\it (i)} The undifferentiated 2-point coincident function,
\begin{eqnarray}
\Bigl\langle \, \Omega \Bigl\vert \,
H_{ij}[A](t, {\vec x}) \, H_{k\ell}[A](t, {\vec x}) 
\, \Bigr\vert \Omega \, \Bigr\rangle  
\nonumber \\
& \mbox{} & \hspace{-5.5cm} 
= 2 \int \frac{d^{D-1}k}{(2\pi)^{D-1}} \;
\Bigl\Vert u[A](t,k) \Bigr\Vert^2 \,
\left[ \, \Pi_{i(k} \Pi_{\ell),j} -
\frac{\Pi_{ij} \Pi_{k\ell}}{D-2} \, \right] 
\;\; , \\
\nonumber \\
& \mbox{} & \hspace{-5.5cm} 
= \frac{2 D (D \!-\! 3)}{(D \!+\! 1) (D \!-\! 2)} \, 
\left[ \, \delta_{i(k} \delta_{\ell)j} - 
\frac{\delta_{ij} \delta_{k\ell}}{D-1} \, \right] \times 
\int \frac{d^{D-1}k}{(2\pi)^{D-1}} \;
\Bigl\Vert u[A](t,k) \Bigr\Vert^2 
\;\; , \qquad \\
\nonumber \\
& \mbox{} & \hspace{-5.5cm} 
\equiv \frac{2 D (D \!-\! 3)}{(D \!+\! 1) (D \!-\! 2)} 
\left[ \, \delta_{i(k} \delta_{\ell)j} - 
\frac{\delta_{ij} \delta_{k\ell}}{D-1} \, \right] \times 
\Bigl\langle \varphi^2 \Bigr\rangle 
\;\; . \label{VEV1}
\end{eqnarray}
{\it (ii)} The ``temporal derivatives'' coincident 2-point function,
\begin{eqnarray}
\Bigl\langle \, \Omega \Bigl\vert \,
\dot{H}_{ij}[A](t, {\vec x}) \, \dot{H}_{k\ell}[A](t,{\vec x}) 
\, \Bigr\vert \Omega \, \Bigr\rangle 
\label{VEV2} \\
& \mbox{} & \hspace{-2.5cm}
= \, \frac{2 D (D \!-\! 3)}{(D \!+\! 1) (D \!-\! 2)} 
\left[ \, \delta_{i(k} \delta_{\ell)j} - 
\frac{\delta_{ij}\delta_{k\ell}}{D - 1} \, \right] 
\times \Bigl\langle \dot{\varphi}^2 \Bigr\rangle 
\;\; . \qquad \nonumber
\end{eqnarray}
{\it (iii)} The ``spatial derivatives'' coincident 2-point function,
\begin{eqnarray}
\Bigl\langle \, \Omega \Bigl\vert \,
\partial_m H_{ij}[A](t, {\vec x}) \, 
\partial_n H_{k\ell}[A](t,{\vec x}) 
\, \Bigr\vert \Omega \, \Bigr\rangle 
\label{VEV3} \\
& \mbox{} & \hspace{-6.5cm}
= \, 
\frac{2 D}{(D \!+\! 3) (D \!+\! 1) (D \!-\! 1) (D \!-\! 2)}
\Bigl[ \, -(D \!+\! 1) \delta_{ij} \delta_{k\ell} \delta_{mn}
+ 4 \delta_{ij} \delta_{k(m} \delta_{n)\ell} 
\nonumber \\
& \mbox{} & \hspace{-7.3cm}
+ \, 4 \delta_{k\ell} \delta_{i(m} \delta_{n)j}
+ (D^2 \!-\! 5) \delta_{i(k} \delta_{\ell)j} \delta_{mn} 
- 4(D \!-\! 1) \delta_{i)(k} \delta_{\ell)(m} \delta_{n)(j} 
\, \Bigr] \times 
\Bigl\langle \Vert \vec{\nabla} \varphi \Vert^2  \Bigr\rangle 
\;\; . \quad \nonumber
\end{eqnarray}

\subsection{Nonperturbative Formulation}

The fictitious scalar expectation values in expressions 
(\ref{VEV1}-\ref{VEV3}) can be written as one dimensional integrals 
involving the mode functions $u[A](t,k)$ for general $A(t)$:
\begin{eqnarray}
\Bigl\langle \varphi^2  \Bigr\rangle 
& = & 
\frac1{2^{D-2} \, \pi^{(\frac{D-1}{2})} \, \Gamma(\frac{D-1}{2})} 
\int_H^{\infty} dk \, k^{D-2} \, 
\Bigl\Vert u[A](t,k) \Bigr\Vert^2 
\;\; , \label{undif} \\
\Bigl\langle \dot{\varphi}^2  \Bigr\rangle 
& = & 
\frac1{2^{D-2} \, \pi^{(\frac{D-1}{2})} \, \Gamma(\frac{D-1}{2})} 
\int_H^{\infty} dk \, k^{D-2} \, 
\Bigl\Vert \dot{u}[A](t,k) \Bigr\Vert^2 
\;\; , \\
\Bigl\langle \Vert \vec{\nabla} \varphi \Vert^2  \Bigr\rangle 
& = & 
\frac1{2^{D-2} \, \pi^{(\frac{D-1}{2})} \, \Gamma(\frac{D-1}{2})} 
\int_H^{\infty} dk \, k^D \, 
\Bigl\Vert u[A](t,k) \Bigr\Vert^2 
\;\; .
\end{eqnarray}
Some work remains to be done converting the expression for $u[A](t,k)$ 
obtained in \cite{TW9,Tomi} to a form for which these integrals are useful,
but this seems straightforward. In the remainder of this section we
explain how (\ref{undif}) can be used to compute leading log expectation 
values of nonlinear functions of the undifferentiated free field 
$H_{ij}[A](t,\vec{x})$ such as those in the nonperturbative evolution
equation (\ref{nperteqn}).

Let us begin with the expectation value of some analytic function $F(x)$
of the fictitious scalar $\varphi(t,\vec{x})$. Because the function is 
analytic we have:
\begin{equation}
F(\varphi) \equiv \sum_{n=0}^{\infty} \frac{F^{n}(0) \varphi^n}{n!} 
\;\; .
\end{equation}
Because the fictitious scalar is free we have:
\begin{equation}
\Bigl\langle \varphi^{2N-1} \Bigr\rangle = 0 
\qquad , \qquad
\Bigl\langle \varphi^{2N} \Bigr\rangle = 
(2N\!-\!1)!! \times \Bigl\langle \varphi^2 \Bigr\rangle^N  
\;\; .
\end{equation}
Thus, the expectation value of $F(\varphi)$ is:
\begin{equation}
\Bigl\langle F(\varphi) \Bigr\rangle = 
\sum_{n=0}^{\infty} \frac{F^{2n}(0)}{2^n n!} 
\times \Bigl\langle \varphi^2 \Bigr\rangle^n 
\;\; .
\end{equation}
In view of the identity:
\begin{equation}
\frac1{{\sqrt \pi}} 
\int_{-\infty}^{\infty} dy \; e^{-y^2} y^{2n} 
\, = \,
\frac{(2n-1)!!}{2^n}
\;\; , \label{yint}
\end{equation}
we reach the final form:
\begin{equation}
\Bigl\langle F(\varphi) \Bigr\rangle = 
\frac{1}{\sqrt{\pi}}
\int_{-\infty}^{\infty} dy \; e^{-y^2} \;
F \Bigl( y \, \sqrt{2 \langle \varphi^2 \rangle} \, \Bigr) 
\;\; . \label{strick}
\end{equation}

We seek now the expectation value of an analytic function of the
matrix of free fields $H_{ij}[A](t,\vec{x})$. Since they are free,
any such expectation value will break up into sums of products of
expectation values of pairs of these fields, possibly with different 
indices. Expression (\ref{VEV1}) allows us to reduce the expectation 
value of any $H_{ij}$ pair to a tensor constant times 
$\langle \varphi^2 \rangle$. There are many integral representations
analogous to (\ref{strick}) for correctly representing the combinatoric
factors. Rather than integrating over a traceless dummy matrix, we 
choose the simpler option of integrating over a symmetric dummy 
variable $Q_{ij}$ and then enforcing tracelessness directly:
\begin{eqnarray}
\lefteqn{\Bigl\langle F(H_{ij}) \Bigr\rangle = } 
\nonumber \\
& & \hspace{-0.3cm}
\frac1{\pi^{(\frac{D^2-D}{4})}} 
\int [dQ_{ij}] \, e^{-Q_{ij} Q_{ij}} \,
F\Biggl( \Bigl[ Q_{ij} - \frac{\delta_{ij} \, Q_{kk}}{D\!-\!1} \Bigr] 
\sqrt{ \frac{4 D (D\!-\!3)}{(D\!+\!1)
(D\!-\!2)} \, \Bigl\langle \varphi^2 \Bigr\rangle } \, \Biggr) 
\;\; . \qquad
\end{eqnarray}

There is no need to continue working in $D$ dimensions since we are only 
interested in the ultraviolet finite, leading logarithm contributions 
from such expectation values. We can therefore take $D=4$:
\begin{equation}
\Bigl\langle \!\! \Bigl\langle F(H_{ij}) 
\Bigr\rangle \!\! \Bigr\rangle \; = \; 
\frac1{\pi^3} \int [dQ_{ij}] \, e^{-Q_{ij} Q_{ij}} \, 
F\Biggl( \Bigl[ Q_{ij} - \frac13 \delta_{ij} \, Q_{kk} \Bigr] 
{\sqrt{ \frac85 \; 
\Bigl\langle \!\! \Bigl\langle \varphi^2 
\Bigr\rangle \!\! \Bigr\rangle}} \, \Biggr) 
\;\; . \label{F(g)}
\end{equation}
It is also very convenient to transform $Q$ to a diagonal form 
${\cal D}$ via a 3-dimensional rotation ${\cal R}$:
\begin{eqnarray}
{\cal R}_{ij} \!\!& \equiv &\!\! 
\Bigl( \delta_{ij} - \widehat{\theta}_i \widehat{\theta}_j \Bigr) 
\cos(\theta) + 
\epsilon_{ijk} \, \widehat{\theta}_k \, \sin(\theta) + 
\widehat{\theta}_i \, \widehat{\theta}_j 
\;\; , \\
Q \!\!& = &\!\! 
{\cal R} \, {\cal D} \, {\cal R}^{T}
\qquad , \qquad
Q_{ij} \; = \; {\cal R}_{ik} \, {\cal D}_{k\ell} \, {\cal R}_{j\ell} 
\;\; , \label{D}
\end{eqnarray}
This is advantageous because the rotation resides only on free indices 
in complicated functions such as the inverse metric and the 4-index
object $B[M]_{ijk\ell}$ of expression (\ref{B}):
\begin{eqnarray}
\exp \Bigl[ -\kappa Q \!+\! \frac13 {\rm Tr}(\kappa Q) \, I \Bigr]_{ij} 
\!\!\! & = & \!\!
{\cal R}_{im} {\cal R}_{jn} \times 
\exp \Bigl[ -\kappa {\cal D} \!+\! 
\frac13 {\rm Tr}(\kappa {\cal D}) \, I \Bigr]_{mn} 
\;\; , \\
B \Bigl[ \kappa Q \!-\! \frac13 {\rm Tr}(\kappa Q) \, I \Bigr]_{ijk\ell} 
\!\!\! & = & \!\!
{\cal R}_{i m} {\cal R}_{j n} {\cal R}_{k p} {\cal R}_{\ell q} \times
B \Bigl[ \kappa {\cal D} \!-\! \frac13 {\rm Tr}(\kappa {\cal D}) \, 
I \Bigr]_{mnpq} 
\qquad
\end{eqnarray}
and drops entirely out of the trace of $Q_{ij}$ in the exponential:
\begin{equation}
Q_{ij} \, Q_{ij} = 
{\cal D}_{11}^2 + {\cal D}_{22}^2 + {\cal D}_{33}^2 
\; \equiv \; 
{\cal D}_{1}^2 + {\cal D}_{2}^2 + {\cal D}_{3}^2
\;\; . \label{trQ}
\end{equation} 

The measure factor in (\ref{F(g)}) takes the form:
\begin{equation}
\int d^6 Q \; = \; 
\int d^3 {\cal R} \int d^3 {\cal D} \; J
\;\; , \label{QtoRD}
\end{equation}
where $J$ is the Jacobian determinant of the $6 \times 6$ matrix of
derivatives of the six $Q_{ij}$'s with respect to the six variables 
$(\theta^i,{\cal D}_j)$.  Because $J$ is independent of $\theta^i$ 
we can compute it using the small angle approximation:
\begin{equation}
{\cal R}_{ij} \; = \; 
\delta_{ij} \, + \, \epsilon_{ijk} \theta_k \, + \,
O(\theta^2)
\;\; . \label{smallangle}
\end{equation}
Then the transformation (\ref{D}) gives:
\begin{eqnarray}
Q_{ii} 
\!\!& = &\!\! 
{\cal D}_i \, + \, O(\theta^2)
\;\; , \nonumber \\
Q_{12} = Q_{21} 
\!\!& = &\!\!
-({\cal D}_1 - {\cal D}_2) \, \theta_3 \, + \, O(\theta^3)
\;\; , \nonumber \\
Q_{23} = Q_{32} 
\!\!& = &\!\!
-({\cal D}_2 - {\cal D}_3) \, \theta_1 \, + \, O(\theta^3)
\;\; , \nonumber \\
Q_{31} = Q_{13} 
\!\!& = &\!\! 
-({\cal D}_3 - {\cal D}_1) \, \theta_2 \, + \, O(\theta^3)
\;\; .
\end{eqnarray}
The Jacobian determinant is therefore:
\begin{equation}
J \; = \; \Bigl\vert ({\cal D}_1 - {\cal D}_2) ({\cal D}_2 - {\cal D}_3)
({\cal D}_3 - {\cal D}_1) \Bigr\vert
\;\; . \label{J}
\end{equation}

Integrating out the angular variables gives a constant $N$:
\begin{eqnarray}
\int \, [ dQ_{ij} ] \; e^{- Q_{ij} Q_{ij}} 
& = &
\nonumber \\
& \mbox{} & \hspace{-2.5cm}
N \int d^3{\cal D} \;
\vert ({\cal D}_1 - {\cal D}_2) ({\cal D}_2 - {\cal D}_3)
({\cal D}_3 - {\cal D}_1) \vert \,\,
e^{- ( \, {\cal D}_1^2 + {\cal D}_2^2 + {\cal D}_3^2 \, )}
\;\; . \label{measure}
\end{eqnarray}
We can find $N$ by transforming to average and relative variables:
\begin{equation}
{\bar {\cal D}} \equiv ( {\cal D}_1 + {\cal D}_2 + {\cal D}_3 )
\quad ; \quad
\Delta_1 \equiv ( {\cal D}_1 - {\cal D}_2 )
\quad , \quad
\Delta_2 \equiv ( {\cal D}_2 - {\cal D}_3 )
\;\; , \label{cm}
\end{equation}
so that:
\begin{eqnarray}
\int d^3{\cal D} \;
\Bigl\vert ({\cal D}_1 - {\cal D}_2) ({\cal D}_2 - {\cal D}_3)
({\cal D}_3 - {\cal D}_1) \Bigr\vert \,\,
e^{- ( \, {\cal D}_1^2 + {\cal D}_2^2 + {\cal D}_3^2 \, )}
\nonumber \\
& \mbox{} & \hspace{-10cm}
=
\int_{-\infty}^{\infty} d{\bar {\cal D}} \; e^{-3 {\bar {\cal D}}^2}
\int_{-\infty}^{\infty} d\Delta_1 
\int_{-\infty}^{\infty} d\Delta_2 \;
\Bigl\vert \Delta_1 \, \Delta_2 ( \Delta_1 + \Delta_2 ) \Bigr\vert \;
e^{- \frac23 ( \, \Delta_1^2 + \Delta_1 \, \Delta_2 + \Delta_2^2 \, )}
\qquad \nonumber \\    
& \mbox{} & \hspace{-10cm}
=
{\sqrt \frac{\pi}{3}}
\int_{-\infty}^{\infty} d\Delta_1 
\int_{-\infty}^{\infty} d\Delta_2 \;
\Bigl\vert \Delta_1 \, \Delta_2 ( \Delta_1 + \Delta_2 ) \Bigr\vert \;
e^{- \frac23 ( \, \Delta_1^2 + \Delta_1 \, \Delta_2 + \Delta_2^2 \, )}
\;\; . \label{cm2}
\end{eqnarray}
We then change to sum and difference variables:
\begin{equation}
\alpha \equiv \Delta_1 + \Delta_2
\qquad , \qquad
\beta \equiv \Delta_1 - \Delta_2
\;\; , \label{lightcone}
\end{equation}
and evaluate the trivial Gaussian integrals:
\begin{eqnarray}
\lefteqn{\int d^3{\cal D} \;
\Bigl\vert ({\cal D}_1 - {\cal D}_2) ({\cal D}_2 - {\cal D}_3)
({\cal D}_3 - {\cal D}_1) \Bigr\vert \,\,
e^{- ( \, {\cal D}_1^2 + {\cal D}_2^2 + {\cal D}_3^2 \, )} }
\nonumber \\
& & \hspace{1cm} =
{\sqrt \frac{\pi}{3}} \; \frac12
\int_{-\infty}^{\infty} d\alpha 
\int_{-\infty}^{\infty} d\beta \;
\frac14 \, \Bigl\vert ( \alpha^2 - \beta^2 ) \alpha \Bigr\vert \;
e^{- \frac12 \alpha^2 - \frac16 \beta^2} = 
\frac{3\pi}{\sqrt 8} 
\;\; . \qquad \label{lightcone2}
\end{eqnarray}
Equation (\ref{lightcone2}) implies that the constant $N$ in 
(\ref{measure}) is:
\begin{equation}
N = \frac{\sqrt{8}}{3\pi}
\;\; .
\end{equation}

\subsection{The First Loop}

It is a simple matter to evaluate the right hand side of (\ref{nperteqn}) 
at one loop -- that is, at order $\kappa^2$. Only the zeroth order parts 
of the inverse metric and the 4-index symbol (\ref{B}) can contribute:
\begin{equation}
{\bar g}^{mn} 
\; \longrightarrow \; \delta^{mn} 
\qquad , \qquad 
B_{ijk\ell}[\kappa H] 
\; \longrightarrow \;
\delta_{ik} \delta_{\ell j} 
\;\; .
\end{equation}
Taking account of the fact that $H_{ij}[A](t,\vec{x})$ is transverse
and using expressions (\ref{VEV2}-\ref{VEV3}) gives:
\begin{eqnarray}
\lefteqn{\frac12 (D\!-\!2) 
\left[ (D\!-\!1) \frac{\dot{A}^2}{A^2} - 
\Lambda - \delta \Lambda_1 \kappa^2 H^4 \right] } 
\nonumber \\
& &  
= \, \frac18 \Bigl\langle \, \Omega \Bigl\vert 
\, \kappa^2 \dot{H}_{ij} \, \dot{H}_{ij} \, 
\Bigr\vert \Omega \, \Bigr\rangle 
+ \frac3{8 A^2} \Bigl\langle \, \Omega \Bigl\vert 
\, \kappa^2 H_{ij, m} H_{ij , m} \,
\Bigr\vert \Omega \, \Bigr\rangle 
+ O(\kappa^4) 
\;\; , \qquad \\
& & 
= \, \frac{D (D\!-\!3)}{8} \Biggl\{
\Bigl\langle \kappa^2 \dot{\varphi}^2 \Bigr\rangle 
+ \frac{3}{A^2} \Bigl\langle 
\kappa^2 \Vert \vec{\nabla} \varphi \Vert^2 
\Bigr\rangle \Biggr\}
+ O(\kappa^4) 
\;\; . \label{1loop}
\end{eqnarray}

Expectation values of the fictitious scalar depend upon the
effective scale factor $A(t)$, which is itself a series in $\kappa^2$.
However, at this order we require only the well known de Sitter limit
\cite{phi4}:
\begin{equation}
\Bigl\langle 
\kappa^2 \partial_{\mu} \varphi \, \partial_{\nu} \varphi
\Bigr\rangle_{A = a} = 
-\frac{H^{D-2}}{(4\pi)^{\frac{D}2}} \,
\frac{\Gamma(D)}{\Gamma(\frac{D}2 \!+\! 1)} \, 
\frac12 \kappa^2 H^2 \,
g_{\mu\nu}^{\rm dS} 
\;\; , \label{dmdn}
\end{equation}
where:
\begin{equation}
g_{\mu\nu}^{\rm dS} dx^{\mu} dx^{\nu} 
\; \equiv \;
-dt^2 + a^2(t) d\vec{x} \cdot d\vec{x}
\;\; .
\end{equation}
Substituting in (\ref{1loop}) gives:
\begin{eqnarray}
\frac{D \!-\! 2}2 
\left[ (D \!-\! 1) \frac{\dot{A}^2}{A^2} - 
\Lambda - \delta \Lambda_1 \kappa^2 H^4 \right]
\nonumber \\
& \mbox{} & \hspace{-4.3cm}
= \, \frac{\kappa^2 H^D}{(4\pi)^{\frac{D}2}} \, 
\frac{\Gamma(D \!+\! 1)}{\Gamma(\frac{D}2 \!+\! 1)} \, 
\frac{(4 \!-\! 3D)(D \!-\! 3)}{16} 
+ O(\kappa^4)
\;\; . \qquad 
\end{eqnarray}
Since all factors in the above equation are finite, we can also 
set $D=4$:
\begin{equation}
3 \frac{\dot{A}^2}{A^2} - 3 H^2 - \delta \Lambda_1 \kappa^2 H^4 =
-\frac{3 \kappa^2 H^4}{8 \pi^2} + O(\kappa^4) 
\;\; . \label{1loop2}
\end{equation}
Furthermore, we express (\ref{1loop2}) more simply as an equation for 
the square of $\dot{A}/A$, and we can more fully anticipate the form 
of the next correction:
\begin{equation}
\frac{\dot{A}^2}{A^2} \; = \; 
H^2 \Biggl\{ 1 + \frac13 \delta \Lambda_1 \kappa^2 H^2
-\frac{\kappa^2 H^2}{8 \pi^2} 
+ O\Bigl(\kappa^4 H^4 \ln(a)\Bigr) \Biggr\}
\;\; . \label{final1L}
\end{equation}

At this point we should comment on various other 1-loop computations.
Had we chosen $\delta \Lambda_1$ to null the 1-loop correction to 
$(\dot{A}/A)^2$ the result would be:
\begin{equation}
\Bigl(\delta \Lambda_1 \Bigr)_{\rm synchronous} = 
+ \frac{3}{8 \pi^2} 
\;\; .
\label{dL1}
\end{equation}
A previous computation in a covariant gauge found \cite{TW10}:
\begin{equation}
\Bigl(\delta \Lambda_1 \Bigr)_{\rm covariant} = 
- \frac{3}{2 \pi^2} 
\;\; .
\label{dL2}
\end{equation}
Finelli, Marozzi, Venturi and Vacca worked in synchronous gauge but 
they used an adiabatic regularization to get \cite{FMVV}:
\begin{equation}
\Bigl(\delta \Lambda_1 \Bigr)_{\rm FMVV} = 
- \frac{361}{1920 \pi^2} 
\;\; .
\label{dL3}
\end{equation}
These differences reflect the well-known fact that counterterms can 
depend upon the choice of gauge and upon the regularization technique. 
The physically significant fact is the universal agreement that any 
1-loop shift in $(\dot{A}/A)^2$ can be absorbed into $\delta \Lambda_1$.

One can also estimate the result from just infrared gravitons. A very 
early computation by Ford got \cite{Ford}:
\begin{equation}
\Bigl(\delta \Lambda_1 \Bigr)_{\rm Ford} = 
+ \frac1{\pi^2} 
\;\; . \label{dL4}
\end{equation}
By instead modeling each graviton polarization as a massless, minimally
coupled scalar and restricting to the infrared, we find \cite{TW11}:
\begin{equation}
\Bigl(\delta \Lambda_1 \Bigr)_{\rm infrared} = 
- \frac1{16 \pi^2} 
\;\; . \label{dL5}
\end{equation}
The reason (\ref{dL1}) and (\ref{dL5}) differ in sign highlights an 
important aspect of the leading logarithm approximation. Expression 
(\ref{1loop}) reveals that the 1-loop effect consists of a sum of the 
expectation values of the squares of two operators. If one makes an 
infrared truncation of the resulting mode sums this must give a positive 
effect, which could only be nulled by a negative value of 
$\delta \Lambda_1$. However, that is not the correct way to work at 
leading logarithm order. It is {\it always} the full quantum field 
theory which defines results; the restriction to infrared modes is only 
justified when extracting an infrared logarithm. The differentiated 
operators that appear in expression (\ref{1loop}) cannot produce 
infrared logarithms. What they give instead is constants, and the
values of these constants derive from the ultraviolet as well as the
infrared. The correct procedure is to compute the expectation value from 
the full mode sum, with the regularization in place. When this is done one 
can see from expression (\ref{dmdn}) that the expectation value of 
$\Vert \vec{\nabla} \varphi \Vert^2$ is actually {\it negative}, and 
large enough that it predominates over the positive expectation value 
of $\dot{\varphi}^2$. Of course there is no mystery about the expectation 
value of a square giving a negative result; the automatic subtraction of 
dimensional regularization has subsumed the positive, power law divergence 
-- which does not, in any case, contribute as vacuum energy -- and left 
a finite, negative remainder.

Whether or not we make the choice (\ref{dL1}) to null the 1-loop 
contribution to $\dot{A}/A$, it is still constant. This allows us to 
evaluate the derivatives of expression (\ref{1loop}) to higher order
using the de Sitter result (\ref{dmdn}) with $H$ replaced by $\dot{A}/A$:
\begin{eqnarray} 
\Bigl\langle \kappa^2 \dot{\varphi}^2 \Bigr\rangle 
& = & 
+\frac{3 \kappa^2}{32 \pi^2} \, \frac{\dot{A}^4}{A^4} 
\, + \, O(\kappa^6) 
\;\; , \\
\frac3{A^2} \Bigl\langle 
\kappa^2 \Vert \vec{\nabla} \varphi \Vert^2 
\Bigr\rangle 
& = & 
-\frac{27 \kappa^2}{32 \pi^2} \, \frac{\dot{A}^4}{A^4} 
\, + \, O(\kappa^6) 
\;\; .
\end{eqnarray}
Since the 1-loop correction to $\dot{A}/A$ is constant we can actually
ignore it compared to the leading logarithm terms that derive from the
undifferentiated factors of $\kappa H_{ij}[A](t,\vec{x})$ in the inverse 
metric and in the 4-index symbol $B[\kappa H]_{ijk\ell}$:
\begin{eqnarray} 
\Bigl\langle \!\! \Bigl\langle
\, \kappa^2 \dot{\varphi}^2 \,
\Bigr\rangle \!\! \Bigr\rangle 
& = & 
+ \frac{3 \kappa^2 H^4}{32 \pi^2} 
\, + \, O(\kappa^6) 
\;\; , \label{dt} \\
\frac3{A^2} \, \Bigl\langle \!\! \Bigl\langle
\, \kappa^2 \Vert \vec{\nabla} \varphi \Vert^2 \,
\Bigr\rangle \!\! \Bigr\rangle 
& = & 
- \frac{27 \kappa^2 H^4}{32 \pi^2} 
\, + \, O(\kappa^6) 
\;\; .
\end{eqnarray}

\subsection{The Second Loop}

We are now ready to evaluate expression (\ref{nperteqn}) at 2-loop
order. A useful identity is the expansion of two contracted 4-index 
symbols:
\begin{eqnarray}
\lefteqn{
\Bigl\langle \!\! \Bigl\langle \, \Omega \Bigl\vert 
\, B[\kappa H]_{ijk\ell} \, B[\kappa H]_{jipq} \,
\Bigr\vert \Omega \, \Bigr\rangle \!\! \Bigr\rangle }
\nonumber \\
& & \hspace{-0.4cm}
= \, 
\delta_{qk} \, \delta_{\ell p} 
+ \frac{\kappa^2}{12} \, 
\Bigl\langle \!\! \Bigl\langle \, \Omega \Bigl\vert 
\, \delta_{qk} (H^2)_{\ell p} - 2 H_{q k} \, H_{\ell p} 
+ (H^2)_{q k} \, \delta_{\ell p} \,
\Bigr\vert \Omega \, \Bigr\rangle \!\! \Bigr\rangle 
+ O(\kappa^4) 
\;\; , \qquad \\
& & \hspace{-0.4cm}
= \,
\delta_{qk} \, \delta_{\ell p} + 
\frac1{15} \Bigl[ \, 4\delta_{qk} \, \delta_{\ell p} 
- \delta_{q\ell} \, \delta_{k p} - \delta_{q p} \, \delta_{k \ell}
\, \Bigr] \,
\Bigl\langle \!\! \Bigl\langle
\kappa^2 \varphi^2 
\Bigr\rangle \!\! \Bigr\rangle 
+ O(\kappa^4) 
\;\; . \label{ID1}
\end{eqnarray}
Contracting (\ref{ID1}) into (\ref{VEV2}) and then making use of 
(\ref{dt}) gives the first of the three expansions we require:
\begin{eqnarray}
\lefteqn{
\frac18 \, 
\Bigl\langle \!\! \Bigl\langle \, \Omega \Bigl\vert 
\, B[\kappa H]_{ijk\ell} \, B[\kappa H]_{jipq} \,
\Bigr\vert \Omega \, \Bigr\rangle \!\! \Bigr\rangle
\times 
\Bigl\langle \, \Omega \Bigl\vert 
\, \kappa^2 \dot{H}_{k\ell} \, \dot{H}_{pq} \, 
\Bigr\vert \Omega \, \Bigr\rangle } 
\nonumber \\
& & \hspace{1.1cm} 
= \, 
\Biggl\{ \frac12 + \frac1{10} 
\Bigl\langle \!\! \Bigl\langle
\, \kappa^2 \varphi^2 \,
\Bigr\rangle \!\! \Bigr\rangle
+ O(\kappa^4) \Biggr\} 
\times
\Bigl\langle \kappa^2 \dot{\varphi}^2 \Bigr\rangle 
\,\; , \\
& & \hspace{1.1cm} 
= \, 
\Biggl\{ \frac12 + 
\frac1{10} \times \frac{\kappa^2 H^2}{ 4 \pi^2} \, \ln(a) 
+ O\Bigl(\kappa^4 H^4 \ln^2(a) \Bigr) \Biggr\} 
\times
\frac{3 \kappa^2 H^4}{32 \pi^2} 
\;\; . \qquad 
\label{Exp1'}
\end{eqnarray}

It is by now clear that we are doing an expansion in powers of a 
dimensionless time dependent parameter which may as well be named:
\begin{equation}
x(t) \; \equiv \;
\frac{\kappa^2 H^2}{4 \pi^2} \, \ln(a) 
\;\; . \label{xdef}
\end{equation}
It is also clear that every term in our expansion will involve one 
factor of $\dot{x}H$. (Higher derivatives can and do appear in
subdominant logarithm corrections, but they cannot occur at leading
order because every time derivative eliminates a factor of 
$t = \ln(a)/H$. This is another way of seeing that only a renormalization
of the cosmological constant is required at leading logarithm order.) 
With this notation the first of our expansions (\ref{Exp1'}) assumes the 
simple form:
\begin{eqnarray}
\lefteqn{
\frac18 \, 
\Bigl\langle \!\! \Bigl\langle \, \Omega \Bigl\vert 
\, B[\kappa H]_{ijk\ell} \, B[\kappa H]_{jipq} \,
\Bigr\vert \Omega \, \Bigr\rangle \!\! \Bigr\rangle
\times 
\Bigl\langle \, \Omega \Bigl\vert 
\, \kappa^2 \dot{H}_{k\ell} \, \dot{H}_{pq} \, 
\Bigr\vert \Omega \, \Bigr\rangle } 
\nonumber \\
& & \hspace{5cm} 
= \, 
H \dot{x} 
\left\{ \, \frac{3}{16} + \frac{3}{80} \, x + 
O(x^2) \, \right\} 
\;\; . \qquad \label{Exp1}
\end{eqnarray}

The inclusion of an inverse metric in relation (\ref{ID1}) adds 
only a single extra term at the order we are working:
\begin{eqnarray}
\lefteqn{
\Bigl\langle \!\! \Bigl\langle \, \Omega \Bigl\vert 
\, B[\kappa H]_{ijk\ell} \, B[\kappa H]_{jipq} \, {\bar g}^{mn} \,
\Bigr\vert \Omega \, \Bigr\rangle \!\! \Bigr\rangle }
\nonumber \\
& & \hspace{-0.5cm}
= \, 
\delta_{qk} \, \delta_{\ell p} \, \delta_{mn} 
+ \frac{\kappa^2}{2} \,
\Bigl\langle \!\! \Bigl\langle \, \Omega \Bigl\vert 
\, \delta_{qk} \, \delta_{\ell p} \, (H^2)_{mn} \, 
\Bigr\vert \Omega \, \Bigr\rangle \!\! \Bigr\rangle 
\nonumber \\
& & \hspace{0.5cm}
+ \, 
\frac{\kappa^2}{12} \, \delta_{mn} \, 
\Bigl\langle \!\! \Bigl\langle \, \Omega \Bigl\vert 
\, \delta_{qk} (H^2)_{\ell p} - 2 H_{q k} H_{\ell p} + 
(H^2)_{q k} \, \delta_{\ell p} \,
\Bigr\vert \Omega \, \Bigr\rangle \!\! \Bigr\rangle 
+ O(\kappa^4) 
\; , \\
& & \hspace{-0.5cm}
= \, 
\delta_{qk} \, \delta_{\ell p} \, \delta_{mn} 
+ \frac1{15} \Bigl[ \, 
14\delta_{qk} \, \delta_{\ell p} - \delta_{q\ell} \, \delta_{k p} 
- \delta_{q p} \, \delta_{k \ell} \, \Bigr] \, \delta_{mn} 
\Bigl\langle \!\! \Bigl\langle 
\kappa^2 \varphi^2 
\Bigr\rangle \!\! \Bigr\rangle 
+ O(\kappa^4) 
\; , \qquad \\
& & \hspace{-0.5cm}
= \,
\delta_{qk} \, \delta_{\ell p} \, \delta_{mn} 
+ \frac1{15} \Bigl[ \, 
14\delta_{qk} \, \delta_{\ell p} - \delta_{q\ell} \, \delta_{k p} 
- \delta_{q p} \, \delta_{k \ell} \, \Bigr] \, \delta_{mn} \, x  
+ O(x^2) 
\;\; . \label{ID2}
\end{eqnarray}
The overall factor of $\delta_{mn}$ in (\ref{ID2}) gives a simple 
result when contracted into the derivative term (\ref{VEV3}):
\begin{eqnarray}
\lefteqn{
\frac1{A^2} \Bigl\langle \, \Omega \Bigl\vert 
\, \kappa^2 \partial_m H_{k\ell} \, \partial_m H_{pq} \,
\Bigr\vert \Omega \, \Bigr\rangle } 
\nonumber \\
& & \hspace{1.5cm} 
= \frac{2 D (D \!-\! 3)}{(D \!+\! 1) (D \!-\! 2)} 
\left[ \, \delta_{k(p} \delta_{q) \ell} - 
\frac{\delta_{k\ell} \, \delta_{pq}}{D \!-\! 1} \, \right] 
\times 
\frac1{A^2} 
\Bigl\langle 
\kappa^2 \Vert \vec{\nabla} \varphi \Vert^2 \, 
\Bigr\rangle 
\;\; . \qquad 
\end{eqnarray}
We can take the $D=4$ limit of the above equation because it 
is finite:
\begin{eqnarray}
\lefteqn{
\frac1{A^2} \Bigl\langle \, \Omega \Bigl\vert 
\, \kappa^2 \partial_m H_{k\ell} \, \partial_m H_{pq} \,
\Bigr\vert \Omega \, \Bigr\rangle } 
\nonumber \\
& & \hspace{2cm} 
\longrightarrow \;
-\frac{9}{10} H \dot{x} \left[ \, 
\delta_{k(p} \delta_{q) \ell} - 
\frac13 \, \delta_{k\ell} \, \delta_{pq} \, \right] 
+ O(H \dot{x} x) 
\;\; . \label{ID2b}
\end{eqnarray}
Multiplying the two factors (\ref{ID2}, \ref{ID2b}) gives the second of 
the three expansions we require for equation (\ref{nperteqn}):
\begin{eqnarray}
\lefteqn{
\frac38 \,
\Bigl\langle \!\! \Bigl\langle \, \Omega \Bigl\vert 
\, B[\kappa H]_{ijk\ell} \, B[\kappa H]_{jipq} \, {\bar g}^{mn} \,
\Bigr\vert \Omega \, \Bigr\rangle \!\! \Bigr\rangle 
\times 
\frac1{A^2} \Bigl\langle \, \Omega \Bigl\vert 
\, \kappa^2 \partial_m H_{k\ell} \, \partial_n H_{pq} \,
\Bigr\vert \Omega \, \Bigr\rangle } 
\nonumber \\
& & \hspace{5.5cm} 
= \, 
H \dot{x} 
\left\{ -\frac{27}{16} - \frac{117}{80} \, x
+ O(x^2) \right\} 
\;\; . \qquad \label{Exp2}
\end{eqnarray}

The two factors of the final term on the right hand side of 
(\ref{nperteqn}) cannot be so usefully expanded in isolation of one 
another. It is better to keep them together so that transversality 
can be exploited. The result is:
\begin{eqnarray}
\lefteqn{
-\frac12 \Bigl\langle \!\! \Bigl\langle \, \Omega \Bigl\vert 
\, B[\kappa H]_{nik\ell} \, B[\kappa H]_{mjpq} \, {\bar g}^{ij} \,
\Bigr\vert \Omega \, \Bigr\rangle \!\! \Bigr\rangle
\times 
\frac1{A^2} \Bigl\langle \, \Omega \Bigl\vert 
\, \kappa^2 \partial_m H_{k\ell} \, \partial_n H_{pq} \,
\Bigr\vert \Omega \, \Bigr\rangle } 
\nonumber \\
& & \hspace{-0.3cm}
= 
-\frac18 \Bigl\langle \!\! \Bigl\langle \, \Omega \Bigl\vert 
\, \kappa^2 H_{ij} H_{k\ell} \,
\Bigr\vert \Omega \, \Bigr\rangle \!\! \Bigr\rangle
\times 
\frac1{A^2} \Bigl\langle \, \Omega \Bigl\vert 
\, \kappa^2 \partial_{\ell} H_{im} \, \partial_j H_{km} \,
\Bigr\vert \Omega \, \Bigr\rangle 
+ O(\kappa^6) 
\; , \\
& & \hspace{-0.3cm}
= 
-\frac1{10} \left[ \delta_{i (k} \delta_{\ell) j} 
- \frac{\delta_{ij} \delta_{k\ell}}3 \right] x 
\times 
\frac{3}{10} \left[ \delta_{i (j} \delta_{\ell) k} 
- 2 \delta_{ik} \delta_{j\ell} \right] H \dot{x}
+ O(H \dot{x} x^2) 
\; , \qquad \\
& & \hspace{-0.3cm}
= 
H \dot{x} \left\{ 
0 + \frac9{40} \, x + O(x^2) \right\} 
\;\; . \label{Exp3} 
\end{eqnarray}

Substituting expansions (\ref{Exp1}), (\ref{Exp2}) and (\ref{Exp3}) 
in our equation (\ref{nperteqn}) for $\dot{A}/A$ gives:
\begin{equation}
\frac{\dot{A}^2}{A^2} \; = \;
H^2 + \frac13 \delta \Lambda_1 \kappa^2 H^4 
- H \dot{x} \left\{ \,
\frac12 + \frac25 \, x + O(x^2) \right\} 
\;\; .
\end{equation}
Taking the square root and retaining only leading logarithm terms gives:
\begin{equation}
\frac{\dot{A}}{A} \; = \;
H + \frac16 \delta \Lambda_1 \kappa^2 H^3 
- \dot{x} \left\{ \, 
\frac14 + \frac15 \, x + O(x^2) \right\} 
\;\; . \label{Hfinal}
\end{equation}
Integrating reveals corrections to the effective scale factor as a series
in powers of the parameter (\ref{xdef}):
\begin{equation}
\ln[A(t)] \; = \;
\ln(a) + \frac{2 \pi^2}{3} \, \delta \Lambda_1 \, x 
- \frac14 \, x - \frac1{10} \, x^2 + O(x^3) 
\;\; , \label{finalA}
\end{equation}
where we remind ourselves that:
\begin{equation}
x(t) \; \equiv \;
\frac{\kappa^2 H^2}{4 \pi^2} \, \ln(a) 
\;\; . 
\end{equation}

One might think that relation (\ref{finalA}) reflects an unfortunate 
background dependence of our formalism. In the sense that the equation
was derived using perturbation theory it certainly does depend upon
the zeroth order result. However, the larger issue is whether or not
it represents the perturbative expansion of a background independent 
result. For proper consideration of this matter it is crucial to 
distinguish dependence upon the initial state from dependence upon the 
background. The leading logarithm approximation, and our entire physical 
picture, only applies to an initial state which suffers a long period of 
inflation. This is a wide class but not a universal one. Many initial
states are so heavily loaded with gravitons that they suffer gravitational
collapse before beginning to inflate. We have nothing to say about such 
states; our focus is instead on states which are initial empty enough that 
inflation begins. The classical evolution of such a state is to locally
approach the de Sitter geometry we used as the background in (\ref{deS}). 
In that case there is a perfectly background independent way of 
interpreting the factors of $\ln(a) = H t$ in expression (\ref{finalA}):
$H^2$ is the invariant acceleration measured by geodesic deviation and
$t$ is the invariant time from the point of observation to the initial
value surface.

\section{Invariant Acceleration Observable}

Equation (\ref{finalA}) shows that the effective scale factor experiences
secular slowing at 2-loop order. However, this is not enough to conclude
that observers actually experience slowing, quite apart from the validity 
of the Effective Scale Factor Approximation. The problem concerns the
noncommutativity of two operations:
\begin{itemize}
\item{Forming an invariant measure of expansion; and}
\item{Taking the expectation value.}
\end{itemize}
The variable $A(t)$ can be regarded as the one third power of the local
volume factor in synchronous gauge. Had the metric been classical, the
logarithmic time derivative of $A(t)$ would indeed give the expansion
rate. But Unruh has shown that one cannot infer physics this way from the 
expectation value of the metric \cite{Unruh}. Instead of taking the
expectation value and then forming an invariant, the correct procedure is 
form an invariant observable from the quantum metric and then take its 
expectation value. Models of scalar-driven inflation which seemed to
show secular slowing at 1-loop order from the expectation value of the
metric \cite{ABM,AW1} show no such 1-loop effect when examined with an 
invariant expansion operator \cite{AW2,GB}.

Finding an invariant measure of the expansion rate in pure quantum 
gravity is more difficult than for scalar-driven inflation because
one lacks the preferred coordinate system in which the scalar is 
homogeneous. A reasonable proposal is based upon using the equation of 
geodesic deviation to measure the local acceleration between objects 
released from rest at some point $x^{\mu}$ \cite{TW4}. In the synchronous 
gauge we have been using this quantity takes the form:
\begin{equation}
\gamma(x) \; = \; 
\frac{1}{g_{rs} \, \Delta^r \, \Delta^s} \,
\left( \,
\frac12 \, {\ddot g_{ij}} \, - \,
\frac14 \, g^{k\ell} \, {\dot g_{ik}} \, {\dot g_{j\ell}}
\, \right) 
\Delta^i \, \Delta^j
\;\; ,\label{gamma}
\end{equation}
where $\Delta^i$ is the spacelike separation of two initially parallel 
timelike geodesics. 

To make $\gamma(x)$ a full invariant one needs to average over the 
initial separation vector $\Delta^i$ and also the position $x^{\mu}$. 
A $\comp$-number such as $\Delta^i$ cannot transform as a vector because 
only the fields of a quantum field theory transform. We therefore make
use of an old trick \cite{TW12} to convert it into a local Lorentz vector 
using the vierbein field $e^i_{~a}(x)$:
\begin{equation}
\Delta^i 
\; \longrightarrow \;
e^i_{~c}(x) \, \Delta^c 
\;\; .
\end{equation}
It will be seen that this takes $\gamma(x)$ to the form:
\begin{equation}
\gamma(x) \; = \; 
\left( \,
\frac12 \, {\ddot g_{ij}} \, - \,
\frac14 \, g^{kl} \, {\dot g_{ik}} \, {\dot g_{jl}}
\, \right) e^i_{~b} \, e^j_{~c} \, \widehat{n}^b \, \widehat{n}^c
\;\; ,\label{newgamma}
\end{equation}
where $\widehat{n}$ is the $\comp$-number unit vector in the initial
separation direction:
\begin{equation}
\widehat{n}^b \; \equiv \; 
\frac{\Delta^b}{\sqrt{\Delta^c \, \Delta^c}}
\;\; .
\end{equation}
We can now take the $\comp$-number average over directions using
\begin{equation}
\int d^{D-2}\widehat{n} \; \widehat{n}^b \, \widehat{n}^c 
\, = \,
\frac{\delta^{bc}}{D \!-\! 1} 
\;\; ,
\end{equation}
to attain a scalar quantity:
\begin{equation}
\overline{\gamma}(x) \; \equiv \;
\int d^{D-2}\widehat{n} \; \gamma(x) \; = \;
\frac{g^{ij} \ddot{g}_{ij}}{2 (D\!-\!1)} - 
\frac{g^{ij} g^{k\ell} \dot{g}_{ik} \, \dot{g}_{j\ell}}{4 (D \!-\! 1)} 
\;\; . \label{gbar}
\end{equation}

To achieve a full invariant we must multiply by $\sqrt{-g(x)}$ and 
integrate. Because the universe was released in a prepared state at 
$t=0$, we can weight this integral by an arbitrary function of the 
invariant time from the initial value surface. That time is not an 
operator at all in synchronous gauge, it is just the coordinate time 
$t$. Moreover, we may as well choose the weight to be a delta function 
selecting a particular time. That still leaves the integration over 
space at this time. However, because the initial state is homogeneous, 
as is the gauge, we can dispense with even this step, although we do 
still need to multiply by $\sqrt{-g(t,\vec{x})}$. The same procedure 
has already been exploited in computing the 1-loop expectation value 
of an invariant 2-point function in flat space background \cite{TW12}.

We now make the Effective Scale Factor Approximation laid out in relations 
(\ref{metric}-\ref{Aeqn}).\footnote{Note that this entails the potentially
invalid step of treating a part of the average geometry --- the effective
scale factor $A(t)$ --- as though it is the quantum geometry. The fully
correct procedure is to form an invariant operator first and then take
its VEV. Our use of the Effective Scale Factor Approximation will only be
justified if one can derive the approximation at leading logarithm order.
We feel this can be done but we acknowledge that it has not been done yet.}
We first evaluate $\overline{\gamma}$ with the substitution $g_{ij} = A^2 \, 
{\bar g}_{ij}$, then take account of simplifications resulting from the 
tracelessness of $H_{ij}$:
\begin{eqnarray}
\overline{\gamma} 
& \longrightarrow  & 
\frac{\dot{A}^2}{A^2} + 
\frac{d}{dt} \frac{\dot{A}}{A} + 
\frac{\dot{A}}{A} \; \frac{{\bar g}^{ij} \dot{\bar g}_{ij}}{D\!-\!1} + 
\frac{{\bar g}^{ij} \ddot{\bar g}_{ij}}{2 (D\!-\!1)} -
\frac{{\bar g}^{ij} {\bar g}^{k \ell} 
\dot{\bar g}_{ik} \, \dot{\bar g}_{j\ell}}{4 (D\!-\!1)} 
\;\; , \\
& = & 
\frac{\dot{A}^2}{A^2} + 
\frac{d}{dt} \frac{\dot{A}}{A} +
\frac{{\bar g}^{ij} {\bar g}^{k \ell} 
\dot{\bar g}_{ik} \, \dot{\bar g}_{j\ell}}{4 (D\!-\!1)} 
\;\; .
\end{eqnarray}
Let us now take note of the highly significant fact that the measure 
factor is a $\comp$-number in the Effective Scale Factor Approximation:
\begin{equation}
\sqrt{-g(t,\vec{x})} 
\; \longrightarrow \; 
[A(t)]^D 
\;\; .
\end{equation}
We can therefore dispense with it altogether in the expectation value:
\begin{eqnarray}
\lefteqn{
\Bigl\langle \!\! \Bigl\langle \, \Omega \Bigl\vert 
\, \overline{\gamma} \, 
\Bigr\vert \Omega \, \Bigr\rangle \!\! \Bigr\rangle 
\; = \; 
\frac{\dot{A}^2}{A^2} 
+ \frac{d}{dt} \frac{\dot{A}}{A} 
+ \frac1{4(D \!-\! 1)} \,
\Bigl\langle \!\! \Bigl\langle \, \Omega \Bigl\vert 
\, B[\kappa H]_{ijk\ell} \, B[\kappa H]_{jipq} \,
\Bigr\vert \Omega \, \Bigr\rangle \!\! \Bigr\rangle }
\nonumber \\
& & \hspace{7.5cm} 
\times \,
\Bigl\langle \, \Omega \Bigl\vert 
\, \kappa^2 \dot{H}_{k\ell} \dot{H}_{pq} \, 
\Bigr\vert \Omega \, \Bigr\rangle 
\;\; , \qquad \label{gbarVEV}
\end{eqnarray}
where we have also used relation (\ref{dtrel}).

At this stage we substitute equation (\ref{nperteqn}) for the 
effective scale factor in (\ref{gbarVEV}):
\begin{eqnarray}
\lefteqn{
\Bigl\langle \!\! \Bigl\langle \, \Omega \Bigl\vert 
\, \overline{\gamma} \, 
\Bigr\vert \Omega \, \Bigr\rangle \!\! \Bigr\rangle
\; = \;
H^2 + \frac{d}{dt} \frac{\dot{A}}{A} 
+ \frac{\delta \Lambda_1 \kappa^2 H^4}{D\!-\!1} } 
\nonumber \\
& & 
+ \, \frac1{4(D\!-\!2)} 
\Bigl\langle \!\! \Bigl\langle \, \Omega \Bigl\vert 
\, B[\kappa H]_{ijk\ell} \, B[\kappa H]_{jipq} \, 
\Bigr\vert \Omega \, \Bigr\rangle \!\! \Bigr\rangle
\times 
\Bigl\langle \, \Omega \Bigl\vert 
\, \kappa^2 \dot{H}_{k\ell} \dot{H}_{pq} \,
\Bigr\vert \Omega \, \Bigr\rangle 
\nonumber \\
& & 
+ \, \frac1{(D \!-\! 2)(D \!-\! 1)} 
\Biggl\langle \!\!\! \Biggl\langle \Omega \Biggl\vert 
\, \frac34 B[\kappa H]_{ijk\ell} \, B[\kappa H]_{jipq} \; {\bar g}^{mn} 
\nonumber \\
& & \hspace{1cm} 
- \, B[\kappa H]_{nik\ell} \, B[\kappa H]_{mjpq} \; {\bar g}^{ij} \,
\Biggr\vert \Omega \Biggr\rangle \!\!\! \Biggr\rangle
\times 
\frac1{A^2} \Bigl\langle \, \Omega \Bigl\vert 
\, \kappa^2 H_{k\ell , m} \, H_{pq , n} \, 
\Bigr\vert \Omega \, \Bigr\rangle 
\; . \qquad 
\end{eqnarray}
We have encountered each of the three terms on the right hand side
(with slightly different numerical coefficients) in the equation for 
the effective scale factor; taking $D=4$ and substituting expansions 
(\ref{Exp1}), (\ref{Exp2}) and (\ref{Exp3}) gives:
\begin{equation}
\Bigl\langle \!\! \Bigl\langle \, \Omega \Bigl\vert 
\, \overline{\gamma} \, 
\Bigr\vert \Omega \, \Bigr\rangle \!\! \Bigr\rangle 
\; = \;
H^2 + \frac{d}{dt} \frac{\dot{A}}{A} +
\frac13 \delta \Lambda_1 \kappa^2 H^4 - 
\frac38 H \dot{x} 
\left\{ \, 1 + x + O(x^2) \, \right\} 
\;\; .
\end{equation}
>From expression (\ref{Hfinal}) we see that the time derivative 
of $\dot{A}/A$ is always subdominant, and we should choose 
$\delta \Lambda_1$ so that initially the observable equals
to $H^2$:
\begin{equation}
\Bigl\langle \!\! \Bigl\langle \, \Omega \Bigl\vert 
\, \overline{\gamma}(t=0) \, 
\Bigr\vert \Omega \, \Bigr\rangle \!\! \Bigr\rangle 
= H^2
\qquad \Longrightarrow \qquad
\delta \Lambda_1 = \frac{9}{32\pi^2} 
\;\; .
\end{equation}
Our final result is:
\begin{equation}
\Bigl\langle \!\! \Bigl\langle \, \Omega \Bigl\vert 
\, \overline{\gamma} \, 
\Bigr\vert \Omega \, \Bigr\rangle \!\! \Bigr\rangle 
\; = \;
H^2 \, \Biggl\{
1 - \frac38\left( \frac{\kappa H}{2 \pi} \right)^4 \, \ln(a) 
+ O\Bigl( \kappa^6 H^6 \ln^2(a) \Bigr) \Biggr\} 
\;\; . \label{gamfin}
\end{equation}
Hence, the observable $\overline{\gamma}$ measures a 
slowdown of the expansion rate.

\section{Epilogue}

During inflation quantum gravitation loop corrections are enhanced
by factors of the number of e-foldings since inflation began. These
enhancement factors are known as {\it infrared logarithms} and there
can be at most one such factor for each extra power of $G H^2$ in
perturbation theory. The set of terms for which this maximum number 
is attained is known as the {\it leading logarithm approximation}.

Quantum gravity is vastly better behaved in the leading logarithm
approximation than in general. The reason for this is that, to
reach leading logarithm order, every undifferentiated free field
must contribute to an infrared logarithm, which precludes these
fields from producing ultraviolet divergences. From the plethora
of BPHZ counterterms which permeate full quantum gravity, only the 
1-loop renormalization of the cosmological constant contributes 
at leading logarithm order.

We have explored a dynamical assumption called the {\it Effective
Scale Factor Approximation}. Under this assumption the graviton remains 
transverse-traceless and free, but propagates in the background
geometry of an effective scale factor $A(t)$ which is determined from
the expectation value of the $g_{00}$ equation. By extending the 
formalism to the first subdominant logarithm order one can even 
describe cosmological perturbations of the correct strength.

Under the Effective Scale Factor Approximation all perturbative
computations are straightforward. We obtain explicit 1- and 2-loop
results for the effective scale factor and for an invariant observable 
based on using the equation of geodesic deviation to probe the local 
acceleration. It is even possible to give a nonperturbative, numerical 
formulation to quantum gravity within the context of the Effective 
Scale Factor Approximation.

The Effective Scale Factor Approximation may not be entirely
correct but it does illustrate how simple inflationary quantum gravity 
can become at leading logarithm order. It also serves as a serious 
test of the ``null hypothesis'' that there are no significant corrections 
from infrared gravitons. For if the hypothesis was correct, then either 
the effective scale factor $A(t)$ would receive no corrections or else 
they would drop out when evaluating the expectation value of the invariant
acceleration observable. However, neither of these possibilities occurs.

\vspace{1cm}

\centerline{\bf Acknowledgements}
This work was partially supported by European Union grant 
MRTN-CT-2004-512194, by Hellenic grant INTERREG IIIA, by 
NSF grant PHY-0653085, and by the Institute for Fundamental 
Theory at the University of Florida.

\end{document}